\documentclass[twocolumn]{aastex631}
\usepackage{threeparttable}
\usepackage{rotating}
\usepackage{xspace}
\usepackage{amsmath}
\usepackage{subfigure}

\newcommand{\flexce}{\texttt{flexCE}\xspace}
\newcommand{\imf}{$\alpha$\xspace}
\newcommand{\ifm}{$M_{\rm inf}$\xspace}
\newcommand{\ift}{$\tau_{\rm inf}$\xspace}
\newcommand{\mini}{$M_{\rm ini}$\xspace}
\newcommand{\sfe}{$\rm SFE$\xspace}

\newcommand{\mgfe}{$\mathrm{[Mg/Fe]}$\xspace}
\newcommand{\feh}{$\mathrm{[Fe/H]}$\xspace}
\newcommand{\afe}{$\mathrm{[\alpha/Fe]}$\xspace}

\begin{document}

\title{Star formation history of Draco revealed by DESI DR1: suppressed early star formation efficiency for Milky Way dwarf galaxies}

\correspondingauthor{Yujiao Yang; Maosheng Xiang }
\email{yujiaoyang7@gmail.com, msxiang@nao.cas.cn }

\author[0000-0002-3180-2327]{Yujiao Yang}
\affiliation{School of Astronomy and Space Science, University of Chinese Academy of Sciences, Beijing 100049, People's Republic of China}

\author[0000-0002-5818-8769]{Maosheng Xiang}
\affiliation{National Astronomical Observatories, Chinese Academy of Sciences, Beijing 100101, People's Republic of China}
\affiliation{ Institute for Frontiers in Astronomy and Astrophysics, Beijing Normal University, Beijing 102206, People's Republic of China}

\author[0000-0003-3250-2876]{Yang Huang}
\affiliation{School of Astronomy and Space Science, University of Chinese Academy of Sciences, Beijing 100049, People's Republic of China}
\affiliation{National Astronomical Observatories, Chinese Academy of Sciences, Beijing 100101, People's Republic of China}

\author[0000-0001-9293-131X]{Meng Zhang}
\affiliation{National Astronomical Observatories, Chinese Academy of Sciences, Beijing 100101, People's Republic of China}
\affiliation{Max-Planck-Institut für Astronomie, Königstuhl 17, D-69117 Heidelberg, Germany}

\author[0009-0004-6855-5149]{Lanya Mou}
\affiliation{National Astronomical Observatories, Chinese Academy of Sciences, Beijing 100101, People's Republic of China}
\affiliation{School of Astronomy and Space Science, University of Chinese Academy of Sciences, Beijing 100049, People's Republic of China}

\author[0000-0002-1802-6917]{Chao Liu}
\affiliation{National Astronomical Observatories, Chinese Academy of Sciences, Beijing 100101, People's Republic of China}
\affiliation{ Institute for Frontiers in Astronomy and Astrophysics, Beijing Normal University, Beijing 102206, People's Republic of China}
\affiliation{University of Chinese Academy of Sciences, Beijing 100049, People's Republic of China}

\author{Jifeng Liu}
\affiliation{National Astronomical Observatories, Chinese Academy of Sciences, Beijing 100101, People's Republic of China}
\affiliation{ Institute for Frontiers in Astronomy and Astrophysics, Beijing Normal University, Beijing 102206, People's Republic of China}
\affiliation{School of Astronomy and Space Science, University of Chinese Academy of Sciences, Beijing 100049, People's Republic of China}
\affiliation{WHU-NAOC Joint Center for Astronomy, Wuhan University, Wuhan, Hubei 430072, People's Republic of China}

\begin{abstract}
Dwarf galaxies serve as fossils for understanding the early star formation history of the local Universe. In this work, we use stellar abundances for individual stars derived from the first data release of the Dark Energy Spectroscopic Instrument (DESI DR1) to investigate the star formation history of the dwarf spheroidal galaxy Draco. We simultaneously fit the stellar metallicity distribution function (MDF) and the [Mg/Fe]–[Fe/H] trend using the galactic chemical evolution (GCE) model \texttt{flexCE}. Our best-fit models imply a low star formation efficiency {\rm SFE} of 0.003--0.007 ~${\rm Gyr}^{-1}$, which is about ten times lower than values reported in literature, but consistent with recent SFE estimates for ultra-faint dwarf galaxies. By combining recent literature measurements, we find that the SFEs of Milky Way dwarf galaxies scale with stellar mass as ${\rm SFE}\propto M_\star^{0.24}$. A low star formation efficiency is associated with a strong outflow efficiency; we estimate a large wind mass-loading factor $\eta_{\rm wind}$ of $\simeq$ 172--260. This is well in line with the observed outflow–stellar mass relation and with the Feedback in Realistic Environment (FIRE) cosmological simulations. Our results demonstrate the power of large-scale, low-to-medium resolution stellar spectroscopic surveys to deliver robust constraints on the early formation history of galaxies in the local Universe.

\end{abstract}


\section{Introduction}\label{sec:intro} 
Dwarf galaxies serve as fundamental laboratories for the archaeological study of galaxy formation and evolution in the local Universe. They are the most numerous class of galaxies and represent a significant fraction of its building blocks. Owing to their proximity, they can be resolved into individual stars with modern instrumentation, enabling detailed analyses on a star-by-star basis.

Extensive observational efforts have been devoted to studying individual stars in nearby dwarf galaxies. Deep photometric surveys conducted with facilities such as the Hubble Space Telescope (HST) have produced high-quality color–magnitude diagrams (CMDs) for large samples of dwarf galaxies, enabling robust reconstruction of their star formation histories \citep[e.g., ][]{2014ApJ...789..147W, 2014ApJ...789..148W, 2024ApJ...977...11Y, 2025ApJS..279...12Z}. Spectroscopic observations, while more observationally demanding, provide critical constraints on stellar ages and detailed chemical abundances, thereby offering deeper insight into the astrophysical processes governing galaxy formation and evolution \citep[e.g.,][]{2009ARA&A..47..371T, 2011ApJ...727...79K, 2023MNRAS.522.5415A, 2023MNRAS.526.5084J, 2026ApJ...998...47S}. To date, most spectroscopic studies of individual stars in dwarf galaxies have relied on medium-resolution data obtained with large-aperture telescopes \citep{2011ApJ...727...79K}. In parallel, high-resolution spectroscopic surveys such as APOGEE \citep{2022ApJS..259...35A} have targeted luminous stars in several dwarf galaxies and their tidal remnants, enabling precise and homogeneous abundance measurements \citep{2022ApJ...928...30L,2026ApJ...998..117S}.

On the other hand, large-scale spectroscopic surveys such as the Dark Energy Spectroscopic Instrument \citep[DESI;][]{2022AJ....164..207D} are providing increasingly deep and systematic coverage of nearby dwarf galaxies, albeit at relatively low spectral resolution. Recent studies have demonstrated that low-resolution spectra, such as those from DESI, can yield reliable chemical abundances for more than ten elements \citep[e.g.,][]{2024ApJS..273...19Z, 2025ApJS..279....5Z}. In near future, the forthcoming China Space Station Telescope \citep[CSST;][]{2011SSPMA..41.1441Z, 2026SCPMA..6939501C} will deliver deep, slitless spectroscopic surveys in the UV–optical range (255–1000 nm) with high spatial resolution over 17,500 $\rm deg^2$ wide field and 400 $\rm deg^2$ deep field, enabling similar analyses for a substantially larger sample of nearby dwarf galaxies. In light of these observational advances, detailed investigations of nearby dwarf galaxies are becoming increasingly prevalent, contributing to a more comprehensive understanding of the formation and evolution of the local Universe.

In this work, we perform chemical evolution modeling to constrain the formation history of the Draco dwarf spheroidal galaxy (dSph) using stellar abundances derived from the DESI DR1 spectra. The Draco dSph is an early-type galaxy with a total stellar mass of $2.9 \times 10^5 M_\odot$, a mean metallicity of $-1.93 \pm 0.01$ dex, and a distance of $76 \pm 6$ kpc \citep{2012AJ....144....4M}. Draco has been the focus of numerous previous studies \citep[e.g.,][]{1998AJ....115.1888S, 2001ApJ...548..592S,2003MNRAS.345...71L,2004MNRAS.351.1338L,2006PASP..118.1361S,2007A&A...468..927L,2009ApJ...701.1053C,2011ApJ...727...79K,2017PASJ...69...76S,2020AA...641A.127R,2025ApJ...983..117H}. By modeling the metallicity distribution function (MDF), \cite{2004MNRAS.351.1338L, 2007A&A...468..927L} inferred a low star formation efficiency (SFE) of 0.03-$0.05 ~\mathrm{Gyr}^{-1}$ and galactic wind with the mass-loading factor $ \eta_{\rm wind} \simeq 4$--6. \cite{2011ApJ...727...79K} derived a short gas infall timescale of $\lesssim 0.42$ Gyr and a star formation duration of $\sim 0.7$ Gyr.

The DESI survey provides nearly complete coverage of the Draco field, obtaining spectra for more than one hundred stars down to a limiting magnitude of $G \sim 21$ in the Gaia $G$ band. We derive stellar elemental abundances from the DESI DR1 spectra using the DD-Payne method, following the approach of \cite{2024ApJS..273...19Z}. Based on this data set, we employ the \flexce model \citep{2017ApJ...835..224A} to reconstruct the formation and chemical enrichment history of Draco in the \mgfe - \feh plane, as well as its MDF.

This paper is organized as follows. In Section~\ref{sec:data}, we describe the selection of Draco member stars and compare our sample with previous studies. Section~\ref{sec:cem} outlines the adopted chemical evolution model and fitting methodology. The results are presented in Section~\ref{sec:results}, where we discuss the implications for the SFE, outflow efficiency $\eta_{\rm wind}$, star formation history (SFH). Finally, we summarize our conclusions in Section~\ref{sec:sum}.


\section{Data}\label{sec:data}
\subsection{DESI DR1}
The Dark Energy Spectroscopic Instrument (DESI; \citealt{2022AJ....164..207D}) is a wide-field spectroscopic survey that has obtained spectra for millions of galaxies, quasars, and stars over a $\sim 14{,}000~\mathrm{deg}^2$ sky area, with a spectral resolution of $R \simeq 2000$--$5000$ and wavelength coverage from $\lambda = 3600$ to $9824$~\AA. The Milky Way Survey (MWS) will observe approximately seven million stars down to a limiting magnitude of $r = 19$ \citep{2023ApJ...947...37C}. \cite{2026OJAp....955260K} present the stellar Value-Added Catalog (VAC) based on DESI Data Release 1 (DR1; \citealt{2026AJ....171..285D}). This release provides stellar parameters, chemical abundances, and radial velocity measurements for more than four million stars, derived using the \texttt{RVSpecFit} (RVS) and \texttt{FERRE} (SP) pipelines \citep{2006ApJ...636..804A, 2019ascl.soft07013K, 2023ApJ...947...37C, 2024MNRAS.533.1012K}. The merged catalog (\texttt{mwsall-pix-iron.fits}) combines the RVS and SP measurements and includes cross-matched Gaia DR3 source identifiers \citep{2023A&A...674A...1G}.

\subsection{Selection member stars of Draco}

\begin{figure*}[!htb]
\centering
\includegraphics[width=0.9\textwidth]{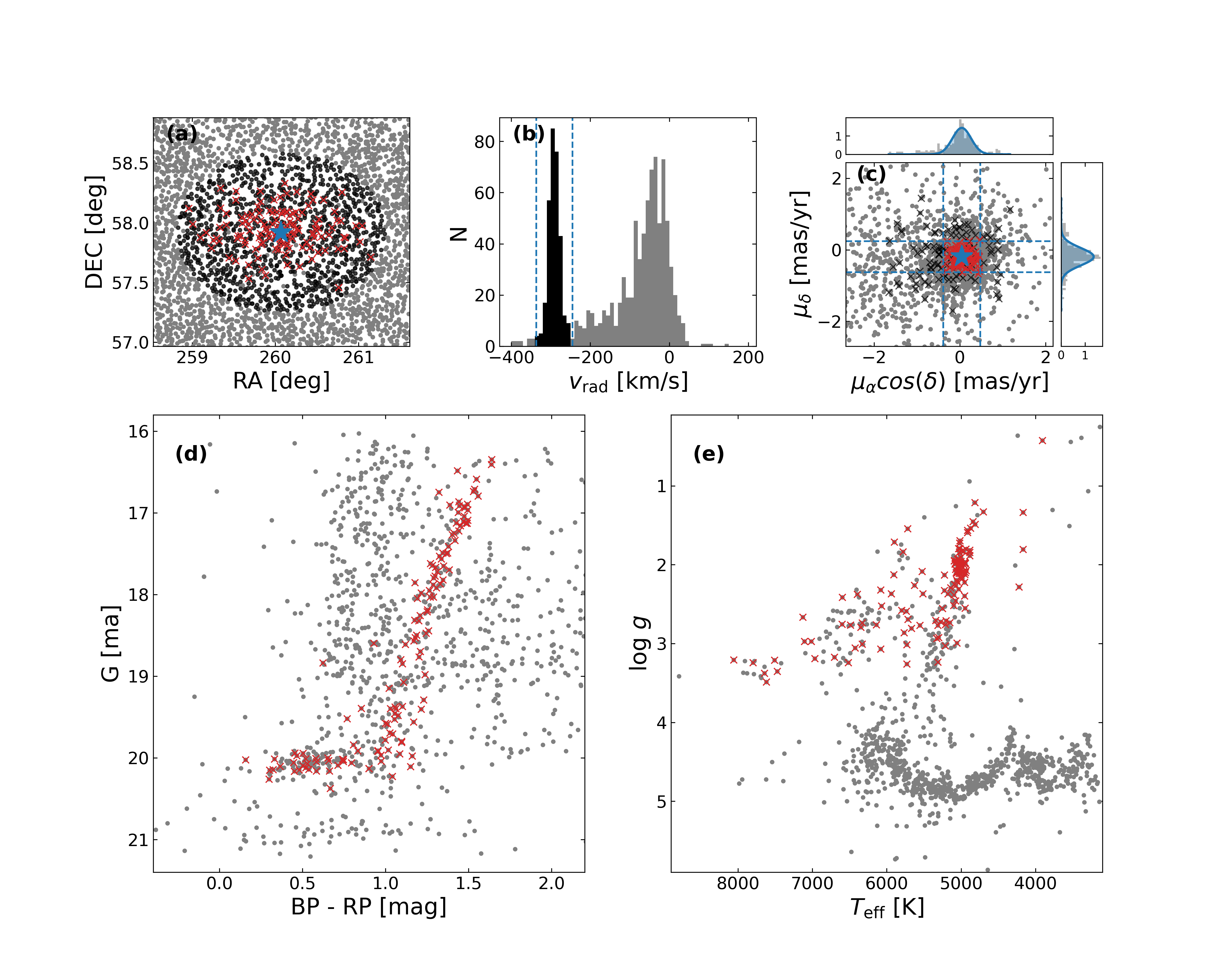}
\caption{Illustrate the process of selecting member stars, the final stars are indicated as red crosses in all panels. Panel (a) : Spatial locations; Panel (b) : Histogram of radial velocities; Panel (c) : proper motion; Panel (d) : Gaia (BP-RP, G) color-magnitude diagram; Panel (e) :  $T_{\rm eff}-\log~g$ diagram. The blue star in Panels (a) and (c) indicate the location center and proper motion center from \cite{2022ApJ...940..136P}.}
\label{fig:select_stars}
\end{figure*}

The member stars of Draco are selected from \texttt{mwsall-pix-iron.fits}. As illustrated in Figure~\ref{fig:select_stars}, the selection follows a few steps as listed below:

\begin{enumerate}
\item Panel (a) :  selecting stars located within the tidal radius \citep[$r_{\rm t} = 48.1^{'}$;][]{2018ApJ...860...66M} of Draco, with the center location of $(\alpha_{\rm 0},\ \delta_{\rm 0})= (260.07^{\circ}, 57.92^{\circ})$ \citep{2022ApJ...940..136P};
\item Panel (b): Selection of stars with line-of-sight velocities within $-336.2 \le v_{\rm los} \le -245.2 \ \mathrm{km\ s^{-1}}$, corresponding to $\pm 5\sigma_{\rm los}$ from the systemic velocity \citep{2022ApJ...940..136P};
\item Panel (c) : selecting stars located within the 2$\sigma$ of the best-fit mean proper motion ($\mu_\alpha cos(\delta)$, $\mu_\delta$). The mean and dispersion ($\sigma$) of the proper motion are deduced with gaussian fit, for $\mu_\alpha cos(\delta)$ and $\mu_\delta$, separately;
\item we further remove Milky Way foreground star contamination using Gaia parallax ($\omega$), with a criterion of $\omega>0.2$~mas. 
\end{enumerate}

These selections result in a sample of 132 member stars. Panel (d) of Figure~\ref{fig:select_stars} shows the member stars' distribution in the Gaia (BP-RP, G) color-magnitude diagram, and Panel (e) show their distribution in the $T_{\rm eff}-\log~g$ diagram, using DESI stellar atmospheric parameters derived with the RVSPecFit pipeline\citep{2024MNRAS.533.1012K}.

\subsection{Stellar abundances}
Based on the DESI early data release \citep[EDR;][]{2024AJ....168...58D}, \citet{2024ApJS..273...19Z} have shown that, stellar abundances for more than 10 individual elements can be robustly derived from the DESI spectra, using the \texttt{DD-Payne}, a physically motivated data-driven machine learning method \citep{2019ApJS..245...34X}. In this work, we derive stellar abundances from DESI DR1 spectra of Draco member stars using a methodology similar to that of \citet{2024ApJS..273...19Z}, but with an updated training set. 

In our implementation, we construct the training set using stars in common between DESI DR1 and the Apache Point Observatory Galactic Evolution Experiment Data Release 17 \citep[APOGEE DR17;][]{2022ApJS..259...35A}, adopting DESI spectra and APOGEE abundances to build a self-consistent spectral model. However, the DESI survey includes a substantial population of very metal-poor stars (${\rm [Fe/H]} < -2$), which are poorly represented in APOGEE, particularly among dwarf stars. To improve coverage in this regime, we additionally include stars in common between DESI DR1 and LAMOST DR9 \citep{2025ApJS..279....5Z} for the training set. Specifically, we select stars with ${\rm [Fe/H]} < -2.3$, $3000 < T_{\rm eff} < 5800~{\rm K}$, and $\log~g < 3.8$, as well as stars with $-2.3 < {\rm [Fe/H]} < -1.0$ and $3500 < T_{\rm eff} < 7500~{\rm K}$.

As shown in Figure~\ref{fig:select_stars}, the Draco member star sample contains a substantial number of blue horizontal branch (BHB) stars, for which \texttt{DD-Payne} abundance estimates are subject to large uncertainties. We therefore exclude these BHB stars from the subsequent analysis, resulting in a final sample of 72 Draco member stars. The full catalog is available in Table~\ref{table:dataall}.

\begin{deluxetable*}{l c c c c c r r r}
\tablecaption{Properties for the Draco Member Stars\label{table:dataall}}
\tabletypesize{\scriptsize}
\setlength{\tabcolsep}{4pt}
\tablewidth{0pt}
\tablehead{
\colhead{Gaia ID} &
\colhead{R.A.} &
\colhead{Decl.} &
\colhead{$V_{\rm rad}$} &
\colhead{$T_{\rm eff}$} &
\colhead{$\log g$} &
\colhead{[Fe/H]} &
\colhead{[Mg/Fe]} &
\colhead{[C/Fe]} \\
\colhead{} &
\colhead{(deg)} &
\colhead{(deg)} &
\colhead{(km s$^{-1}$)} &
\colhead{(K)} &
\colhead{(dex)} &
\colhead{} &
\colhead{} &
\colhead{}
}
\startdata
1432978523635391232 & 261.018455 & 57.841915 & -299.636 & $5124\pm58.1$ & $2.698\pm0.200$ & $-2.175\pm0.080$ & $0.188\pm0.078$ & $0.444\pm0.081$ \\
1432977561562695936 & 260.827632 & 57.783593 & -284.967 & $4905\pm46.2$ & $2.012\pm0.126$ & $-1.500\pm0.065$ & $-0.317\pm0.077$ & $-0.456\pm0.073$ \\
1432977424123741184 & 260.755085 & 57.782224 & -291.070 & $5015\pm59.7$ & $2.428\pm0.183$ & $-2.291\pm0.077$ & $0.363\pm0.090$ & $0.223\pm0.089$ \\
1432931622592641024 & 260.762138 & 57.458133 & -290.198 & $4925\pm31.1$ & $2.324\pm0.102$ & $-2.168\pm0.044$ & $0.186\pm0.046$ & $-0.165\pm0.050$ \\
1433160767687038976 & 260.223897 & 57.911219 & -283.039 & $4821\pm30.1$ & $2.106\pm0.103$ & $-1.979\pm0.041$ & $0.097\pm0.059$ & $-0.263\pm0.058$ \\
1433154892171707392 & 260.056360 & 57.866437 & -291.833 & $4704\pm25.3$ & $1.629\pm0.080$ & $-1.661\pm0.035$ & $-0.155\pm0.047$ & $-0.614\pm0.053$ \\
1433162657472930560 & 260.270127 & 58.033610 & -293.882 & $4810\pm21.4$ & $1.891\pm0.069$ & $-1.616\pm0.028$ & $-0.099\pm0.042$ & $-0.761\pm0.045$ \\
1433162077653154048 & 260.142122 & 57.972901 & -292.660 & $4868\pm43.4$ & $1.807\pm0.134$ & $-1.575\pm0.063$ & $0.016\pm0.078$ & $-0.322\pm0.081$ \\
1433145103942385152 & 260.376147 & 57.837857 & -299.586 & $4597\pm17.7$ & $0.868\pm0.032$ & $-2.117\pm0.023$ & $-0.111\pm0.027$ & $-0.547\pm0.038$ \\
1433161970277971712 & 260.214598 & 57.971696 & -298.565 & $4983\pm64.8$ & $2.497\pm0.234$ & $-2.566\pm0.079$ & $0.525\pm0.084$ & $-0.056\pm0.118$ \\
\enddata
\tablenotetext{}{The first 10 rows are shown here. The full table is available online as a machine-readable table.}
\end{deluxetable*}

\begin{figure*}[!htb]
\includegraphics[width=0.95\textwidth,trim=0.2cm 0.2cm 0.2cm 0.2cm, clip]{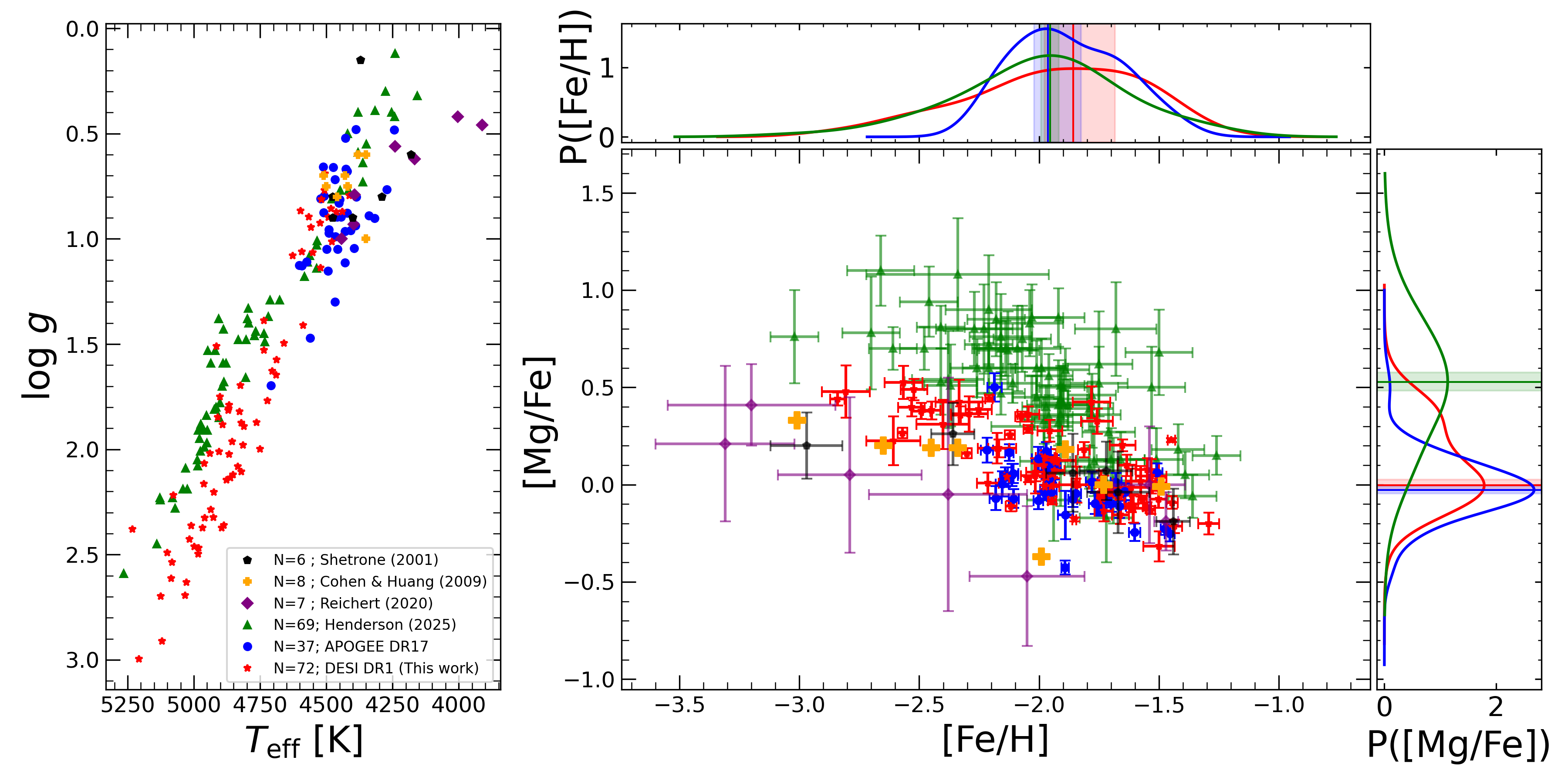}
\caption{Distribution of stellar atmospheric parameters and chemical abundances of Draco member stars from six catalogs. Left panel:  $T_{\rm eff}-\log~g$ diagram;  Right panel: \mgfe--\feh plane. Error bars represent the 1$\sigma$ measurement uncertainties. The top and right panels show the normalized kernel density estimates (KDEs) of the \feh and \mgfe distributions, respectively, convolved with the observational errors. Vertical (horizontal) lines indicate the peak values of the distributions, and the shaded regions represent the $1\sigma$ uncertainties estimated via bootstrap resampling. For the clarity, the top and right panels only include catalogs with number of stars larger than 10. For all panels, the data points are color-coded by their data sources: \cite{2001ApJ...548..592S} (black pentagons), \cite{2009ApJ...701.1053C} (orange pluses), \cite{2020AA...641A.127R} (purple diamonds), \cite{2025ApJ...983..117H} (green triangles), APOGEE DR17 (blue circles), and DESI DR1 (this work; red stars).}
\label{fig:alpha_plots}
\end{figure*}

\begin{deluxetable*}{lcccccrrc}
\tablecaption{Summary of the datasets, including the peak, mean, and range of the \feh\ and \mgfe\ distributions for each sample.\label{table:data}}
\tabletypesize{\scriptsize}
\setlength{\tabcolsep}{4pt}
\tablewidth{0pt}
\tablehead{
\colhead{Catalog} &
\colhead{$N$} &
\colhead{$R$} &
\colhead{$[\mathrm{Fe/H}]_{\rm peak}$} &
\colhead{$[\mathrm{Fe/H}]_{\rm mean}$} &
\colhead{$[\mathrm{Fe/H}]_{\rm range}$} &
\colhead{$[\mathrm{Mg/Fe}]_{\rm peak}$} &
\colhead{$[\mathrm{Mg/Fe}]_{\rm mean}$} &
\colhead{$[\mathrm{Mg/Fe}]_{\rm range}$}
}
\startdata
\cite{2001ApJ...548..592S} & 6  & 34000  & \nodata & $-2.00 \pm 0.51$ & $[-2.97, -1.44]$ & \nodata & $ 0.06 \pm 0.15$ & $[-0.19, 0.26]$ \\
\cite{2009ApJ...701.1053C} & 8  & 35000  & \nodata & $-2.19 \pm 0.48$ & $[-3.01, -1.49]$ & \nodata & $ 0.09 \pm 0.20$ & $[-0.37, 0.33]$ \\
\cite{2020AA...641A.127R}  & 7  & 15000  & \nodata & $-2.39 \pm 0.69$ & $[-3.31, -1.47]$ & \nodata & $-0.01 \pm 0.26$ & $[-0.47, 0.41]$ \\
\cite{2025ApJ...983..117H} & 69 & 7000   & $-1.95^{+0.03}_{-0.04}$ & $-2.00 \pm 0.33$ & $[-3.02, -1.26]$ & $0.53^{+0.05}_{-0.05}$ & $ 0.50 \pm 0.29$ & $[-0.17, 1.10]$ \\
APOGEE DR17               & 37 & 22500  & $-1.96^{+0.12}_{-0.05}$ & $-1.88 \pm 0.21$ & $[-2.20, -1.46]$ & $-0.03^{+0.02}_{-0.02}$ & $-0.02 \pm 0.15$ & $[-0.43, 0.50]$ \\
DESI DR1 (This work)      & 72 & 2000--5000 & $-1.92^{+0.10}_{-0.09}$ & $-1.95 \pm 0.35$ & $[-2.84, -1.29]$ & $-0.00^{+0.03}_{-0.03}$ & $ 0.11 \pm 0.20$ & $[-0.32,  0.53]$ \\
\enddata
\tablenotetext{}{Peak abundances are derived using bootstrap resampling with observational uncertainties incorporated. No peak values are reported for catalogs containing fewer than 10 stars. Mean values are given as the sample mean and standard deviation. The reported ranges correspond to the minimum and maximum abundances in each catalog.}
\end{deluxetable*}

Figure~\ref{fig:alpha_plots} presents the distribution of Draco member stars in the $T_{\rm eff}$--$\log~g$ and \mgfe--\feh planes using stellar parameters and abundances derived with \texttt{DD-Payne}. We compare our measurements with literature samples, including the high-resolution Keck/HIRES studies of \citet{2001ApJ...548..592S} and \citet{2009ApJ...701.1053C}, the compilation of archival high-resolution spectra by \citet{2020AA...641A.127R}, the Keck/DEIMOS medium-resolution analysis of \citet{2025ApJ...983..117H}, and APOGEE DR17. For APOGEE DR17, only stars with \texttt{FLAG} = 0 are retained, resulting in a sample of 37 Draco members.

Figure~\ref{fig:alpha_plots} shows that the DESI \texttt{DD-Payne} [Mg/Fe] measurements are generally consistent with those from APOGEE DR17, as well as with the results of \citet{2001ApJ...548..592S} and \citet{2009ApJ...701.1053C}, in the [Mg/Fe]--[Fe/H] plane. In contrast, the [Mg/Fe] estimates of \citet{2020AA...641A.127R} are systematically lower at fixed metallicity, whereas those of \citet{2025ApJ...983..117H} are systematically higher. These discrepancies likely arise from differences in spectral resolution, stellar atmosphere models, and the treatment of non-local thermodynamic equilibrium (NLTE) effects among the studies. The NLTE effect for [Fe/H] used in this work has been corrected, see details in \cite{2025ApJS..279....5Z}. Table~\ref{table:data} summarizes the [Fe/H] and [Mg/Fe] measurements of Draco member stars, including both our measurements and literature values. We note that the KDE of our [Mg/Fe] distribution exhibits an extended high-[Mg/Fe] tail, while a similarly large dispersion is also seen in other abundance ratios.

\subsection{Identifying archaeology tracers in abundance space}
\begin{figure*}[!htb]
\includegraphics[width=0.95\textwidth,trim=0.2cm 0.2cm 0.2cm 0.2cm, clip]{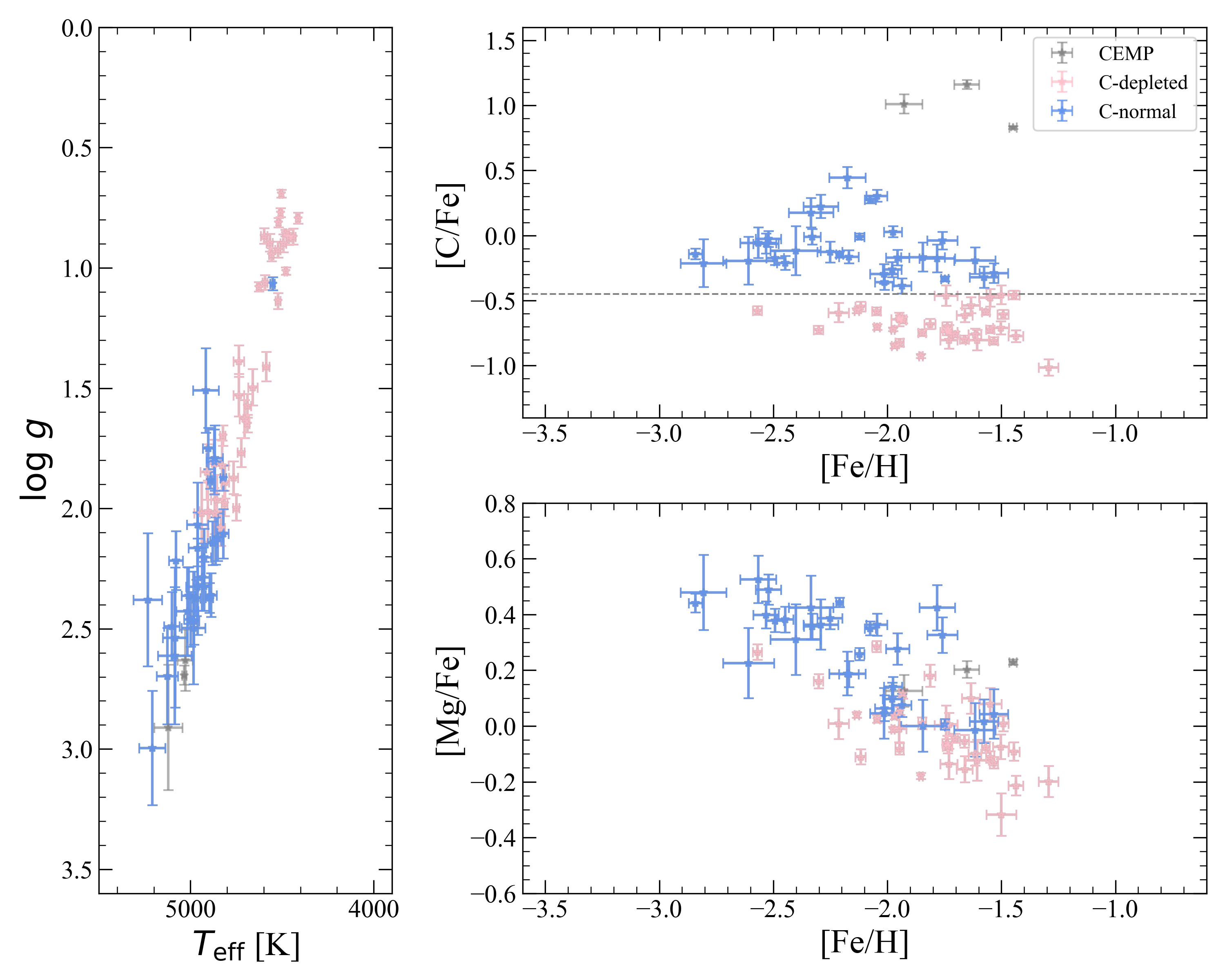}
\caption{Distributions of the CEMP (gray), C-normal (blue), and C-depleted (pink) stars in the $T_{\rm eff}$--$\log g$ diagram (left panel), the [C/Fe]--[Fe/H] plane (upper-right panel), and the [Mg/Fe]--[Fe/H] plane (lower-right panel). The separation between the C-depleted and C-normal populations is determined empirically from the observed [C/Fe] distribution with $\mathrm{[C/Fe]}=-0.45$. }
\label{fig:class}
\end{figure*}

The upper-right panel of Figure~\ref{fig:class} presents the distribution of stars in the [C/Fe]--[Fe/H] plane. The carbon abundances clearly separate the sample into three groups: three stars with $\mathrm{[C/Fe]} \geq 0.7$, which meet the definition of carbon-enhanced metal-poor stars \citep[CEMP;][]{2007ApJ...655..492A}, 37 carbon-depleted (C-depleted) stars, and 32 carbon-normal or slightly enriched stars (C-normal). The division between the C-depleted and C-normal populations is defined empirically from the observed distribution, corresponding to a boundary at $\mathrm{[C/Fe]}=-0.45$. We mention that no correction for evolutionary effects on [C/Fe] has been applied in this work.

In the following analysis, we exclude the three CEMP stars, as their surface abundances may be affected by complex stellar evolution processes, such as third dredge-up and/or binary mass transfer, and therefore may not reliably trace the chemical evolution history of the system. The C-depleted stars are likely tip-RGB objects whose surface [C/Fe] has been depleted by dredge-up during the RGB phase. These stars have, on average, lower $\log g$ (i.e., higher luminosities) than the C-normal stars (left panel of Figure~\ref{fig:class}). The C-normal population likely includes RGB stars that have not experienced significant carbon depletion, as well as asymptotic giant branch (AGB) stars whose surface carbon abundances are either consistent with or slightly enhanced relative to their natal composition.

As shown in the bottom-right panel of Figure~\ref{fig:class}, there is moderate evidence that, at fixed [Fe/H], stars in the C-normal group exhibit slightly higher [Mg/Fe] on average than those in the C-depleted group. Part of this offset may arise from the larger abundance uncertainties at the metal-poor end. Alternatively, some AGB stars within the C-normal population may have experienced evolutionary processes, such as hot-bottom burning and third dredge-up, that modified their surface abundances and potentially enhanced their [Mg/Fe] ratios \citep{2014PASA...31...30K}. If so, these stars may not serve as unbiased tracers of Draco's chemical evolution history.

\section{Chemical evolution modeling}\label{sec:cem}
Galactic chemical evolution (GCE) models have been widely employed to investigate the formation and enrichment histories of the Milky Way and nearby galaxies \citep[e.g.,][]{2009ARA&A..47..371T, 2012ceg..book.....M, 2020ApJ...900..179K, 2021A&ARv..29....5M}. The chemical abundances serve as fossil records of different nucleosynthetic channels. The $\alpha$-elements (O, Mg, Si, S, and Ca) are primarily synthesized in massive stars and released through core-collapse supernovae (SNe II) on timescales of $\sim$1 Myr, whereas a significant fraction of iron-peak elements (Cr, Mn, Fe, Ni, Co, Cu, and Zn) originates from Type Ia supernovae (SNe Ia) on timescales of $\sim$1 Gyr \citep{2020ApJ...900..179K}. In the \afe-\feh plane, $\alpha$-element abundances typically exhibit an initial high plateau due to rapid enrichment by SNe II, followed by a decline as SNe Ia begin to contribute iron. The morphology of the \afe-\feh relation—including the high-$\alpha$ plateau, the slope, the “knee”, and the stellar density distribution along \feh—encodes information about star formation, gas inflow and outflow processes, as well as the detailed SFH.

We use \flexce \citep{2017ApJ...835..224A} to model Draco's chemical enrichment with five parameters: the total mass of inflow gas \ifm, inflow timescale \ift, the wind mass-loading factor $\eta_{\rm wind}$, the star formation efficiency \sfe, and the slope of IMF \imf. \flexce is a flexible one-zone chemical evolution code, assuming that the interstellar medium (ISM) is instantaneously mixed. In our implementation, we assume that the gas inflow follows an exponential function
\begin{equation}
\dot{M}_{\rm inflow} = \frac{M_{\rm inf}}{\tau_{\rm inf}}e^{-t/\tau_{\rm inf}}
\end{equation}
with two fitting parameters \ifm and \ift. The gas outflow is parameterized as 
\begin{equation}
\dot{M}_{\rm outflow} = \eta_{\rm wind} {\rm SFR}
\end{equation}
with one fitting parameter $\eta_{\rm wind}$. The SFR is deduced as 
\begin{equation}
{\rm SFR}  = {\rm SFE}\cdot M_{\rm gas}
\label{func:sfr}
\end{equation}
The stellar mass distribution is assumed to follow a single power law with slope \imf\ over the mass range $0.1$--$100~M_\odot$. All other model parameters are adopted from the default configuration of \citet{2017ApJ...835..224A}. Specifically, the model assumes an initial gas mass of $1\times10^{9}~M_\odot$, core-collapse supernova (CCSN) yields from \citet{2004ApJ...608..405C}, AGB star yields from \citet{2010MNRAS.403.1413K}, Type Ia supernova yields from \citet{1999ApJS..125..439I}, and solar abundances from \citet{2009ARA&A..47..481A}.

We constrain the GCE model by jointly fitting the chemical abundance distribution in the \mgfe-\feh plane and the MDF. The total log-likelihood is defined as
\begin{equation}
    \ln \mathcal{L}(\theta) = \ln \mathcal{L}_{\rm MgFe} +  \omega \ln \mathcal{L}_{\rm MDF}
\end{equation}

where $\theta$ denotes the model parameters, $w=500$ is a scaling factor adopted to balance the statistical weight of the two components. For the abundance trend, we compare the observed $\mathrm{[Mg/Fe]}_{\rm obs}$ against the model predictions $\mathrm{[Mg/Fe]}_{\rm mod}$--interpolated at the observed metallicity \feh -- by assuming a Gaussian likelihood:
\begin{equation}
\begin{split}
    \ln \mathcal{L}_{\rm [Mg/Fe]} = -\frac{1}{2}\sum_{i=1}^{N} \Big[ & \frac{([\rm Mg/Fe]_{obs,\mathit{i}} - [Mg/Fe]_{mod,\mathit{i}})^2}{\sigma_{\mathrm{[Mg/Fe]}_i}^2} \Big]
\end{split}
\end{equation}
where $\sigma_{\mathrm{[Mg/Fe]}_i}$ is the observational uncertainties in [Mg/Fe] of individual stars. Simultaneously, to capture the overall chemical enrichment history, we evaluate the MDF using the Earth Mover’s Distance (EMD) between the observed and theoretical cumulative distribution functions (CDFs). To account for the measurement uncertainties in the observed metallicities, we construct the observed MDF through a Monte Carlo error-propagation procedure. For each star, we generate 500 realizations of its metallicity by perturbing the measured value according to its individual uncertainty. An MDF is constructed for each realization using the same metallicity bins. The mean MDF over the 500 realizations is then adopted as the observed MDF for the likelihood calculation. The model MDF is constructed by binning the stellar mass formed in each metallicity increment, and the distance is calculated as $d_{\rm EMD} = \sum_{k}
\left| C_{\mathrm{mod},k}-C_{\mathrm{obs},k}\right| \Delta x_k$, where $C$ represents the normalized CDF and $\Delta x_{k}$ is the bin width. The corresponding log-likelihood is then formulated as:
\begin{equation}
    \mathrm{ln}~\mathcal{L}_{\rm MDF} = -\frac{1}{\lambda} \sum_{k} \left| C_{\mathrm{mod},k}-C_{\mathrm{obs},k}\right| \Delta x_k
\end{equation}
where $\lambda = 0.05$ is a hyperparameter governing the sensitivity of the MDF fit. This integrated approach ensures that the best-fit model parameters $\theta$ simultaneously reproduce the detailed abundance sequences and the integrated mass distribution of the stellar populations. 

\begin{figure*}[!htb]
\centering
\includegraphics[width=0.9\textwidth,trim=0.2cm 0.2cm 0.2cm 0.2cm,clip]{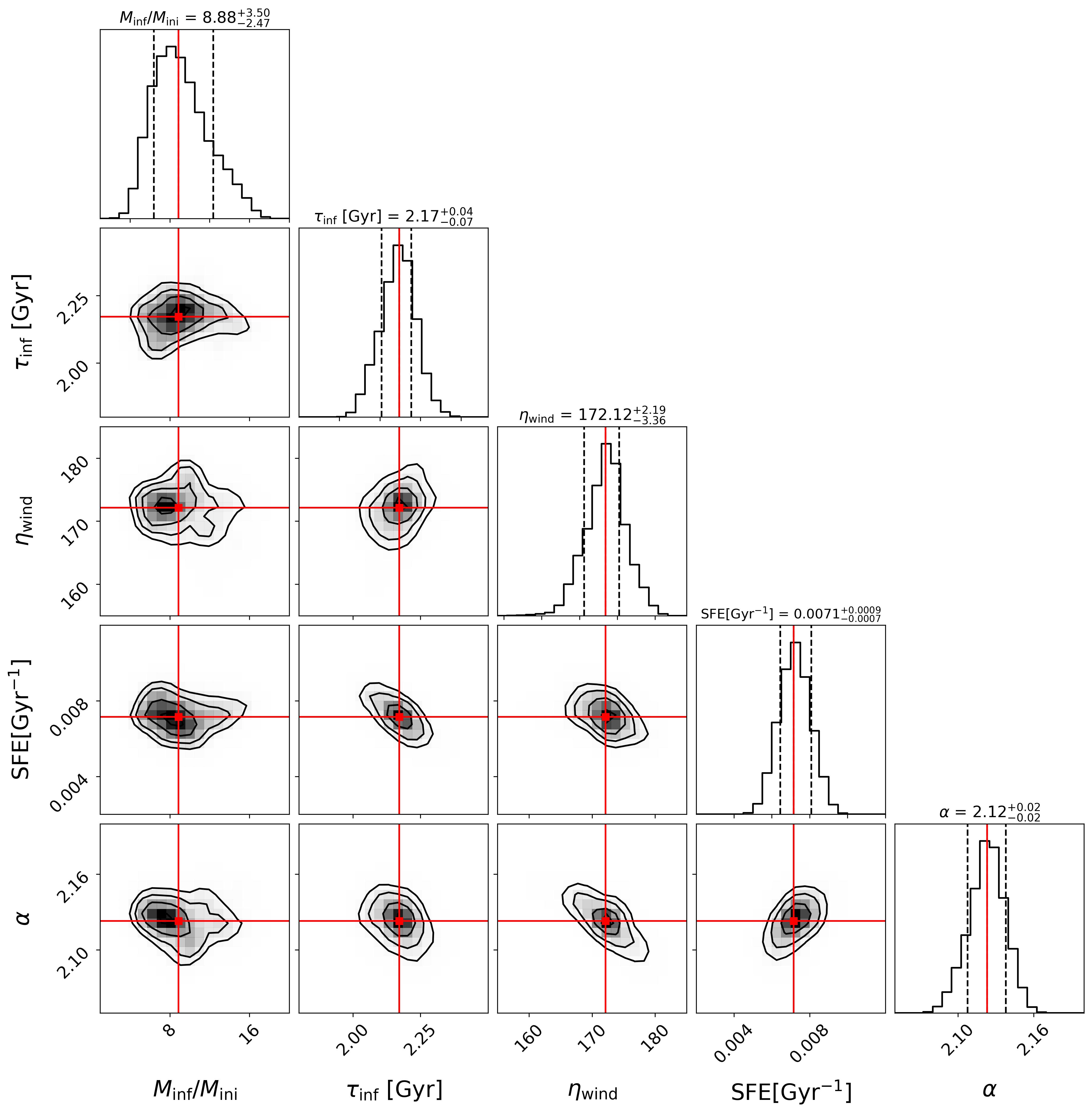}
\caption{Corner plot of the joint posterior distribution for five fitting parameters: the ratio of initial gas mass to inflow gas mass \mini/\ifm, the inflow timescale $\tau_{\rm inf}$ (Gyr), the outflow mass-loading factor $\eta_{\rm wind}$, the star formation efficiency (SFE, Gyr$^{-1}$), and the IMF slope $\alpha$. Diagonal panels show the 1D posterior distributions, with dashed lines marking the 16th–84th percentiles and red lines indicating the median values (also listed in the panel titles). Off-diagonal panels show the 2D posterior contours, with red lines indicating the corresponding median estimates.}
\label{fig:corner_all}
\end{figure*}

The fitting process is completed with the Markov Chain Monte Carlo (MCMC) code \texttt{emcee} \citep{2013PASP..125..306F}. We initialized 50 MCMC walkers over 10000 iterations to derive the best-fit parameters by optimizing the likelihood function.

\section{Results and Discussion}\label{sec:results}

We fit the full sample after excluding CEMP stars, as well as the C-depleted and C-normal stars, individually. A comparison of the results derived from these individual groups provides a self-validation of the chemical modeling.

\begin{figure*}[!htb]
\begin{subfigure}
    \centering
    \includegraphics[width=0.9\textwidth,height=7cm, trim=0.2cm 0.2cm 0.2cm 0.2cm,clip]{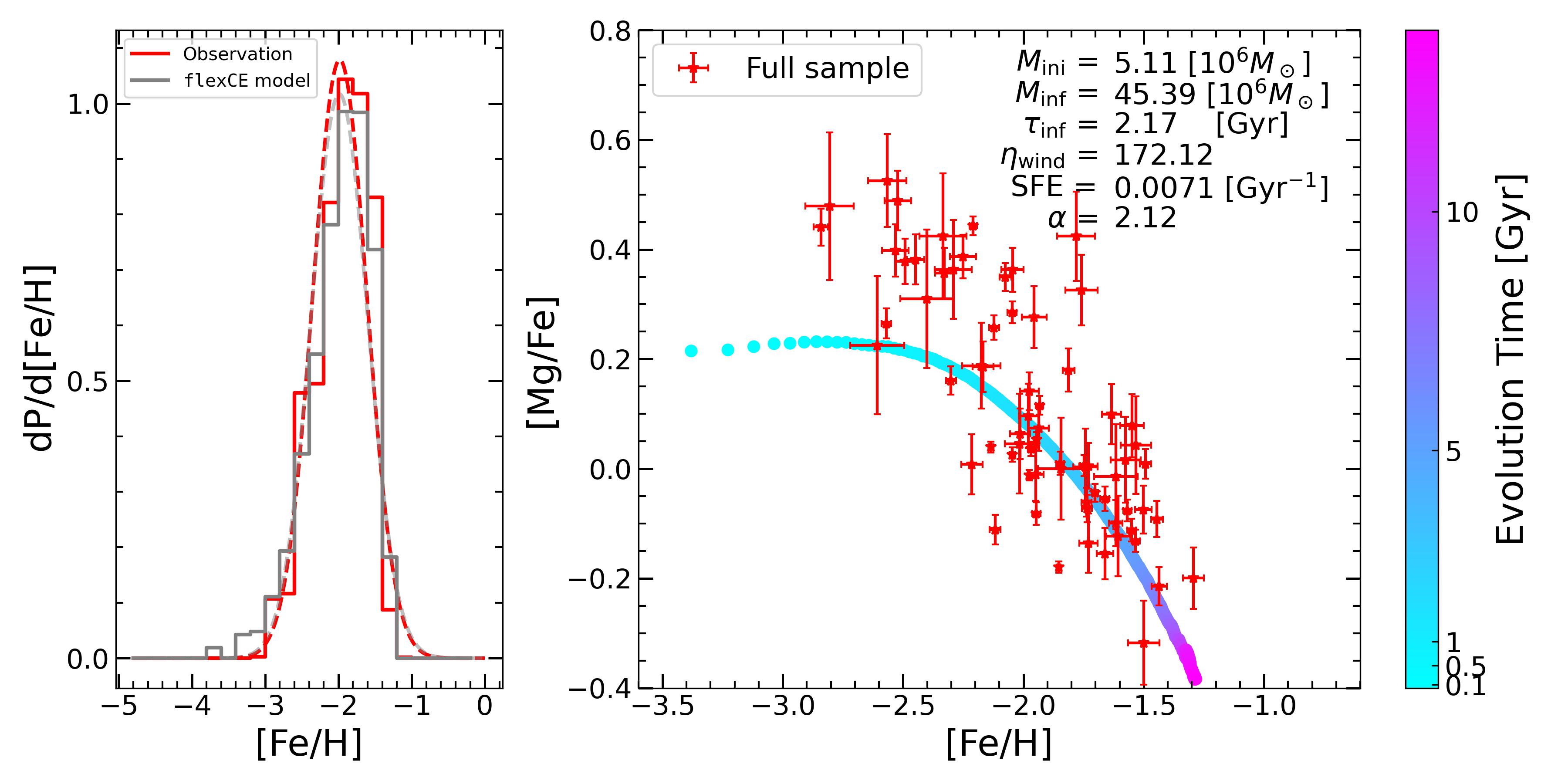}
\end{subfigure}
\begin{subfigure}
    \centering
    \includegraphics[width=0.9\textwidth,height=7cm, trim=0.2cm 0.2cm 0.2cm 0.2cm,clip]{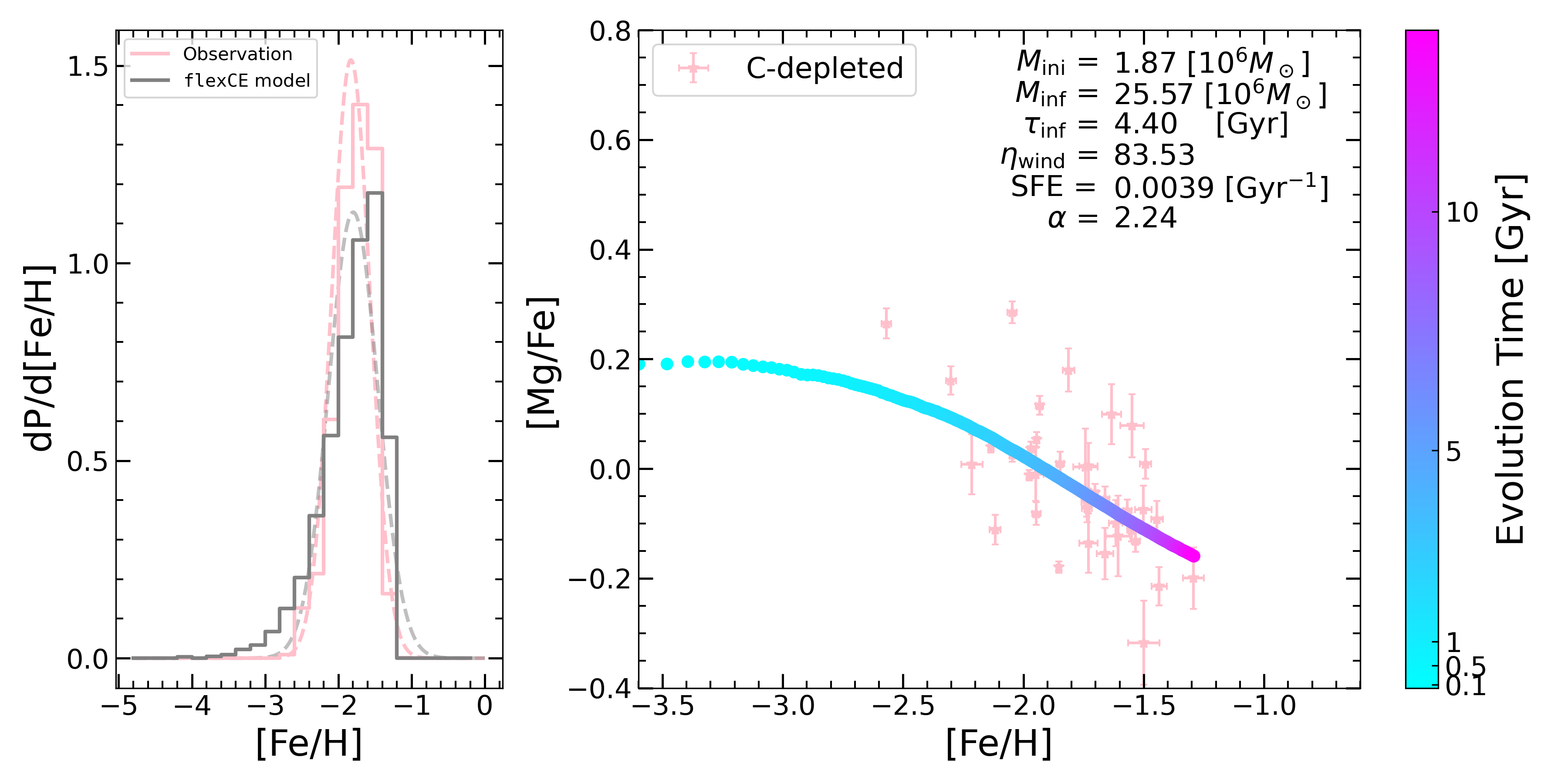}
\end{subfigure}
\begin{subfigure}
    \centering
    \includegraphics[width=0.9\textwidth,height=7cm, trim=0.2cm 0.2cm 0.2cm 0.2cm,clip]{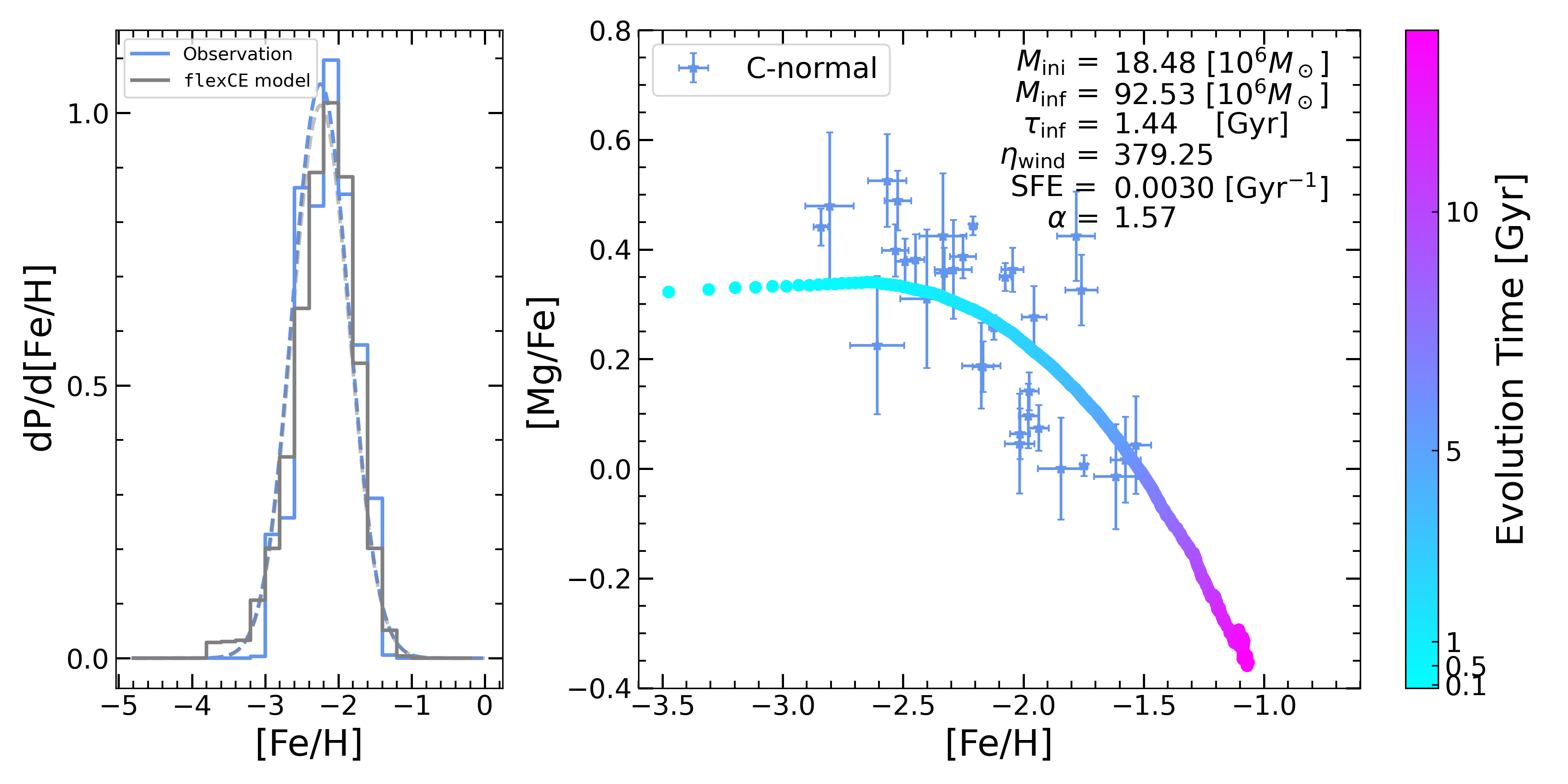}
\end{subfigure}
\vspace{0.5cm}
\caption{Best-fit MDFs (left panels) and \mgfe--\feh relations (right panels) for the full sample (top row), C-depleted stars (middle row), and C-normal stars (bottom row). Left panels show the [Fe/H] distributions with a bin width of 0.1 dex. Observed (modeled) MDFs are shown as colored (gray) solid lines, and dashed lines indicate the best-fit Gaussian approximation. Right panels show the abundance distributions of Draco member stars, with the best-fit GCE model represented by cyan-to-purple points color-coded by evolutionary time. Best-fit parameters are listed in the upper-right corner of each panel. The parameters \mini\ and \ifm\ are scaled to the total stellar mass of $2.9 \times 10^5~M_\odot$ from \citet{2012AJ....144....4M}.}
\label{fig:draco_result}
\end{figure*}

Figure~\ref{fig:corner_all} presents the corner plot of the joint posterior distributions for the best-fit parameters obtained from the full sample.  We note that elemental abundance ratios and gas fractions are independent of the total gas mass \citep{2017ApJ...835..224A}; therefore, the chemical evolution model constrains only relative mass fractions rather than the absolute galaxy mass. For this reason, we parameterize the model in terms of the ratio between the initial gas mass and the total infall gas mass, $M_{\rm ini}/M_{\rm inf}$, which is adopted consistently in all corner plots.

Figure~\ref{fig:draco_result} presents the best-fit GCE models for the full sample (top row), the C-depleted subsample (middle row), and the C-normal subsample (bottom row). The three samples yield moderately different best-fit models, although each reproduces the observed MDF and the [Mg/Fe]--[Fe/H] relation of its corresponding sample reasonably well. The C-depleted model exhibits slower chemical evolution, with the [Mg/Fe] knee occurring at ${\rm [Fe/H]}\lesssim -3$. In contrast, the C-normal population shows more rapid enrichment, with the knee located at ${\rm [Fe/H]}\simeq -2.4$. The full sample fit yields intermediate parameters; however, its best-fit model does not fully reproduce the observed [Mg/Fe] distribution at the metal-poor end. Several inferred parameters differ significantly among the three populations. In particular, the IMF slope is sensitive to the [Mg/Fe] plateau level. The C-depleted sample favors a steeper IMF slope, $\alpha_{\rm IMF}=2.24$, whereas the C-normal population prefers a shallower slope, $\alpha_{\rm IMF}=1.57$, reflecting its systematically higher [Mg/Fe] ratios.

It is not straightforward to determine which population provides the best representation of the underlying GCE model parameters. On the one hand, the magnesium abundances of C-depleted (RGB) stars likely preserve a more pristine fossil record of the early chemical enrichment history. However, the number of stars in this subsample is limited, particularly at the very metal-poor end (${\rm [Fe/H]}<-2$), so the inferred model parameters may be subject to substantial biases and uncertainties. On the other hand, while the C-normal population yields a more robust MDF fit owing to its larger sample size, some of its stars may be evolved AGB stars whose surface magnesium has been altered by stellar evolution processes. As a result, the inferred parameters from this subsample may also be subject to substantial biases and uncertainties. 

For comparison, star-count studies in the literature have derived an IMF slope of $\alpha_{\rm IMF}\simeq 2.1$--2.3 \citep{1998AJ....115..144G}, which is broadly consistent with our estimates for both the full sample and the C-depleted population. However, the star-count method primarily probes the low-mass end of the IMF ($\lesssim 1~M_\odot$), whereas chemical evolution modeling is mainly sensitive to the high-mass regime that dominates nucleosynthetic enrichment. Thus, the agreement is not strictly an independent validation of our inferred IMF slopes.

\subsection{Star formation efficiency}

\begin{figure*}[!hb]
\includegraphics[width=0.95\textwidth,trim=0.1cm 0.1cm 0.1cm 0.1cm,clip]{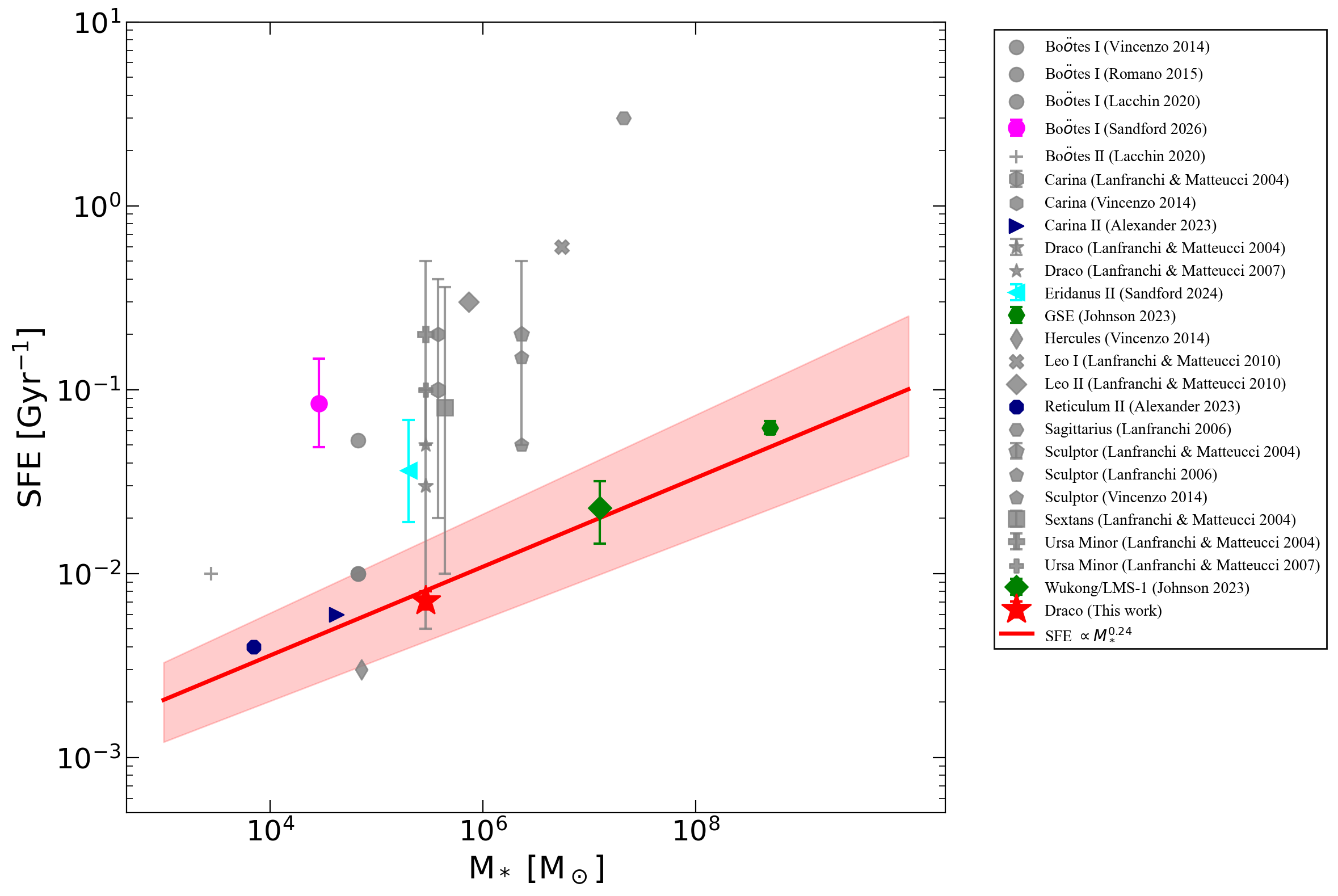}
\caption{Star formation efficiency (SFE) as a function of stellar mass ($M_\star$) for Local Group dwarf galaxies and stellar streams derived from chemical evolution analyses. The red stars denote the Draco results obtained in this work. Error bars indicate reported uncertainties where available. The red solid line shows a power-law relation between SFE and $M_\star$, fitted to systems with quantitatively constrained SFEs from chemical evolution models, including Carina~{\sc ii} and Reticulum~{\sc ii} \citep{2023MNRAS.522.5415A}, GSE and Wukong/LMS-1 \citep{2023MNRAS.526.5084J}, and Draco (this work). Additional literature measurements are taken from \cite{2004MNRAS.351.1338L,2007A&A...468..927L,2010A&A...512A..85L,2014MNRAS.441.2815V,2015MNRAS.446.4220R,2020MNRAS.495.3276L,2024MNRAS.530.2315S,2026ApJ...998...47S}. Stellar masses are adopted from \cite{2021ApJ...909..192G} for Eridanus~{\sc ii}, \cite{2022arXiv220409057N} for Wukong/LMS-1 and GSE, \cite{2017MNRAS.465.1879S} for Carina~{\sc ii} and Reticulum~{\sc ii}, and \cite{2012AJ....144....4M} for the remaining systems.}
\label{fig:sfe_mass}
\end{figure*}

The GCE model places strong constraints on the star formation efficiency (SFE). The full sample and the individual carbon-defined subpopulations all favor low SFEs, with best-fit values ranging from $0.003$ to $0.007~{\rm Gyr}^{-1}$. Adopting a stellar mass of $2.9\times10^{5}~M_\odot$ \citep{2012AJ....144....4M}, we place Draco in the SFE--$M_\star$ diagram in Figure~\ref{fig:sfe_mass}. 

\cite{2004MNRAS.351.1338L} found that the $[\alpha/{\rm Fe}]$ ratios of Draco are best reproduced with a low SFE ($\sim 0.03~{\rm Gyr}^{-1}$) combined with a strong galactic wind (wind efficiency $\omega_{\rm i}=6$). In their subsequent study, \cite{2007A&A...468..927L} incorporated the observed stellar metallicity distribution and reached a similar conclusion, obtaining a slightly higher SFE of $\sim0.05~{\rm Gyr}^{-1}$ and a wind efficiency of $\omega_{\rm i}=4$. They concluded that an efficient and continuous galactic wind is required to reproduce both the metallicity distribution and the present-day gas mass of Draco. However, these previous studies explored the GCE parameter space using relatively coarse grids, with the wind efficiency restricted to a narrow range of $\omega_{\rm i}\sim6$--10. This limited parameter exploration makes it difficult to assess whether the preferred parameters represent the global optimum or to quantify the uncertainties and degeneracies between the SFE and outflow efficiency.

Our best-fit SFE of $\sim0.007~{\rm Gyr}^{-1}$ is lower than those reported by \cite{2004MNRAS.351.1338L, 2007A&A...468..927L}, indicating an even less efficient star formation history in our model. Meanwhile, our inferred $\eta_{\rm wind}\simeq172$ is substantially larger than their wind efficiencies. These values, however, should not be compared directly because the wind prescriptions, parameterizations, and observational constraints differ between the studies. In particular, our $\eta_{\rm wind}$ is an effective, time-averaged parameter constrained simultaneously by the MDF and the $[\mathrm{Mg/Fe}]-[\mathrm{Fe/H}]$ distribution, while the previous studies primarily explored discrete sets of model parameters. Despite these quantitative differences, the overall physical picture is consistent with previous studies: Draco experienced inefficient star formation accompanied by strong gas outflow.

Figure~\ref{fig:sfe_mass} shows that our SFE estimate for Draco is comparable to literature values for ultra-faint dwarf galaxies (UFDs) with stellar masses below $10^{5}~M_\odot$, including Boötes~{\sc i}  and Boötes~{\sc ii} \citep{2020MNRAS.495.3276L}, and Carina~{\sc ii}  and Reticulum~{\sc ii} \citep{2023MNRAS.522.5415A}. Notable exceptions are the higher SFE estimates for Eridanus~{\sc ii} and Boötes~{\sc i} reported by \citet{2024MNRAS.530.2315S} and \citet{2026ApJ...998...47S}, respectively. One possible explanation for these differences is that these studies constrain the GCE models using only the MDF, without simultaneously incorporating detailed abundance patterns.

Our SFE estimate for Draco is consistent with the trend defined by recent measurements from detailed GCE modeling, including Carina~{\sc ii} and Reticulum~{\sc ii} \citep{2023MNRAS.522.5415A}, as well as the stellar populations associated with GSE and Wukong/LMS-1 \citep{2023MNRAS.526.5084J}. In contrast, our inferred SFE for Draco is typically lower by approximately an order of magnitude than previous estimates for dwarf galaxies with stellar masses above $10^5~M_\odot$. These earlier studies \citep{2004MNRAS.351.1338L, 2007A&A...468..927L, 2010A&A...512A..85L, 2014MNRAS.441.2815V, 2015MNRAS.446.4220R, 2020MNRAS.495.3276L} generally adopted relatively simple GCE prescriptions and explored the parameter space using coarse grids with restricted parameter ranges. For example, their outflow mass-loading factors were typically limited to $\eta_{\rm wind}<15$, approximately an order of magnitude below the values inferred from recent GCE studies with broader parameter exploration and more comprehensive chemical constraints (see Section~\ref{subsec:eta}). Such restrictions may limit the ability of the models to capture the degeneracies among the SFE, gas inflow, and outflow efficiency, potentially leading to systematically higher inferred SFEs.

Given that the earlier estimates may be affected by the limitations of the adopted modeling approaches, we suggest that dwarf galaxies in general have lower SFEs than those reported in the earlier studies, approximately following the trend shown in the Figure~\ref{fig:sfe_mass}. This trend was determined by fitting the recent measurements of Carina~{\sc ii} and Reticulum~{\sc ii} \citep{2023MNRAS.522.5415A}, GSE and Wukong/LMS-1 \citep{2023MNRAS.526.5084J}, and Draco (this work). We performed bootstrap resampling with 1000 iterations to estimate the uncertainties in the power-law relation between SFE and stellar mass. In each iteration, we randomly resample the observed data points with replacement and perform an least-squares fit to $\rm log_{10}SFE = \mathit{a}log_{10}M_\star + \mathit{b}$. The distributions of the fitted coefficients were then used to determine the mean values and 16th--84th percentile confidence intervals, which are represented by the pink shaded region in Figure~\ref{fig:sfe_mass}.

\subsection{The outflow efficiency}
\label{subsec:eta}
\begin{figure*}[!htb]
\includegraphics[width=0.95\textwidth,trim=0.1cm 0.1cm 0.1cm 0.1cm,clip]{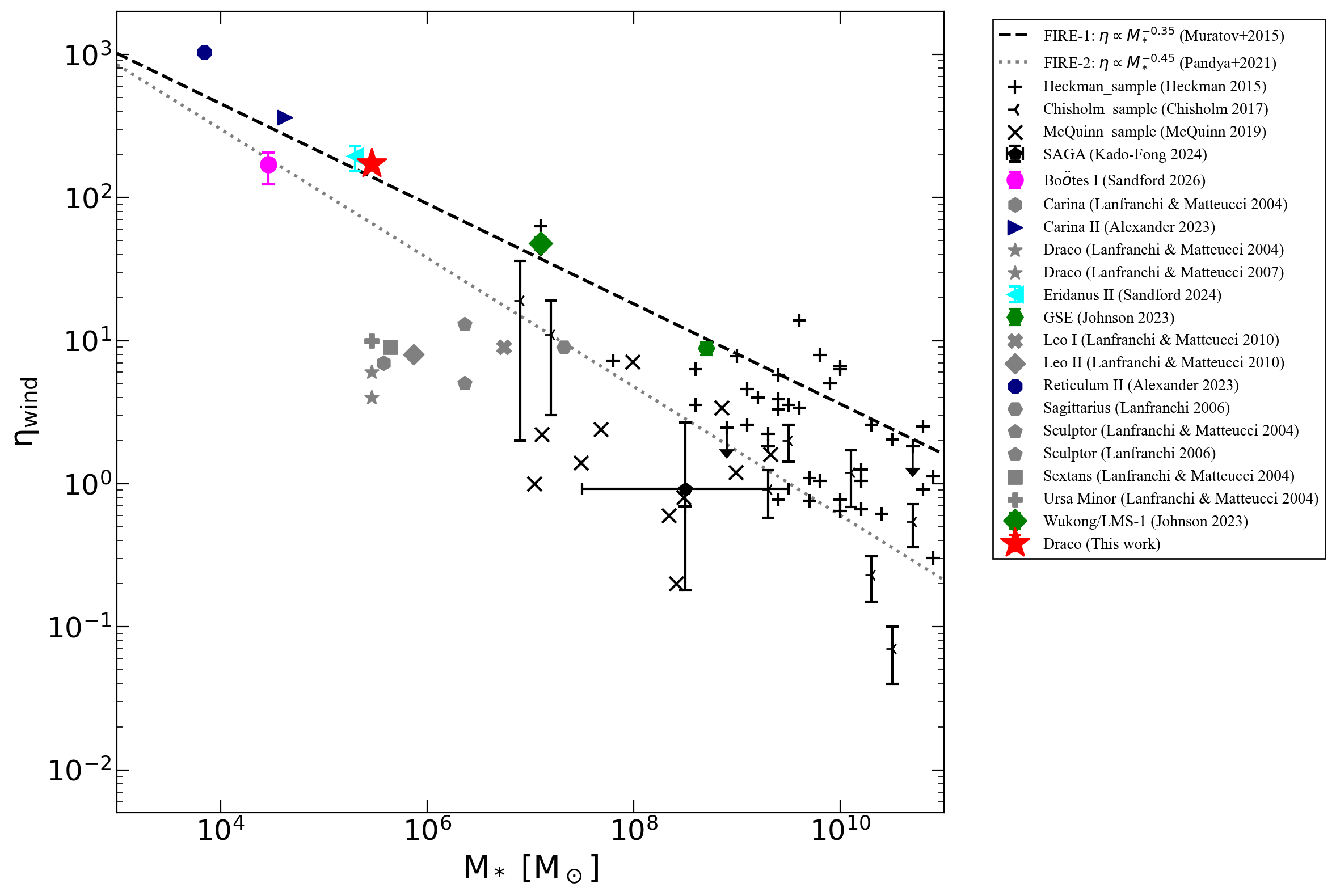}
\caption{Mass-loading factors $\eta_{\rm wind}$ as a function of stellar mass $M_\star$ for dwarf galaxies and stellar streams. GCE-based estimates are shown for classical dwarfs \citep[Carina, Draco, Leo~{\sc i}, Leo~{\sc ii}, Sagittarius, Sculptor, Sextans, Ursa Minor;][]{2004MNRAS.351.1338L, 2006A&A...453...67L, 2007A&A...468..927L, 2010A&A...512A..85L}, ultra-faint dwarf galaxies \citep[Boötes~{\sc i}, Carina~{\sc ii}, Eridanus~{\sc ii},  Reticulum~{\sc ii};][]{2023MNRAS.522.5415A,2024MNRAS.530.2315S,2026ApJ...998...47S} and stellar streams \citep[GSE, Wukong/LMS-1;][]{2023MNRAS.526.5084J}. Mass-loading factors inferred from the evolution of the mass–metallicity relation in the SAGA background dwarf galaxy sample (\citealt{2024ApJ...966..129K}) are also shown, along with direct observational constraints on galactic outflows in nearby star-forming galaxies (\citealt{2015ApJ...809..147H, 2017MNRAS.469.4831C, 2019ApJ...886...74M}). The scaling relations found in cosmological simulations FIRE-1 \citep[$\eta_{\rm wind} = 3.6\times (M_\star/{10^{10} M_\odot})^{-0.35}$;][]{2015MNRAS.454.2691M} and FIRE-2 \citep[$\eta_{\rm wind} = 0.6\times (M_\star/{10^{10} M_\odot})^{-0.45}$;][]{2021MNRAS.508.2979P} are overplotted. For the stellar mass, we adopt the value for Eridanus~{\sc ii} from \cite{2021ApJ...909..192G}, those for Wukong/LMS-1 and GSE from \cite{2022arXiv220409057N}, the masses for Carina~{\sc ii} and Reticulum~{\sc ii} from \cite{2017MNRAS.465.1879S}, and for the remaining galaxies from \cite{2012AJ....144....4M}.}
\label{fig:eta_mass}
\end{figure*}

GCE modeling applied to the full sample yields an outflow mass-loading factor of $172.12^{+2.19}_{-3.36}$. This parameter shows moderate variations among the individual subpopulations: the C-depleted population gives a value of $83.53^{+2.98}_{-2.79}$, while the C-normal population gives $379.25^{+3.88}_{-3.45}$. Same as Figure 12 in \cite{2024MNRAS.530.2315S}, Figure~\ref{fig:eta_mass} presents the outflow mass-loading factor ($\eta_{\rm wind}$) as a function of galactic stellar mass ($M_\star$), compiling measurements from GCE models for dwarf galaxies \citep[Carina, Draco, Leo~{\sc i}, Leo~{\sc ii}, Sagittarius, Sculptor, Sextans, Ursa Minor;][]{2004MNRAS.351.1338L, 2006A&A...453...67L, 2007A&A...468..927L, 2010A&A...512A..85L}, ultra-faint dwarf galaxies \citep[Boötes~{\sc i}, Carina~{\sc ii}, Eridanus~{\sc ii},  Reticulum~{\sc ii};][]{2023MNRAS.522.5415A,2024MNRAS.530.2315S,2026ApJ...998...47S} and stellar streams \citep[GSE, Wukong/LMS-1;][]{2023MNRAS.526.5084J}. Direct measurements of galactic outflows in nearby star-forming galaxies \citep{2015ApJ...809..147H, 2017MNRAS.469.4831C, 2019ApJ...886...74M} and scaling relation from cosmological simulations Feedback in Realistic Environment \citep[FIRE-1;][]{2015MNRAS.454.2691M} and FIRE-2 \citep{2021MNRAS.508.2979P} also presented. 

The figure shows that our measurements for Draco align well with the overall trend defined by literature values, except for the series of measurements by \citep{2004MNRAS.351.1338L, 2006A&A...453...67L,2007A&A...468..927L,2010A&A...512A..85L}. As noted above, these latter measurements yield values around 10 because the authors imposed an upper limit of 15 on the outflow mass-loading factor in their modeling.

We emphasize that the mass-loading factors shown in Figure~\ref{fig:eta_mass} are not strictly equivalent quantities. In our GCE model,  $\eta_{\rm wind}$ is an effective, time-averaged parameter that describes the cumulative impact of outflows on the chemical evolution and is constrained by the observed MDF and [Mg/Fe]-[Fe/H] distribution. In cosmological simulations such as FIRE, the mass-loading factor is generally defined as the instantaneous outflow rate relative to the SFR, measured at a specified radius (e.g., 0.25 $R_{\rm vir}$) and averaged over selected redshift intervals. For a galaxy with the same stellar mass as Draco, the FIRE-1 and FIRE-2 simulations yield $\eta_{\rm wind}\simeq 139$ and $\eta_{\rm wind}\simeq 66$, respectively. Similarly, observations of nearby star-forming galaxies generally trace the instantaneous or recent outflow state through UV absorption lines \citep[e.g.,][]{2015ApJ...809..147H, 2017MNRAS.469.4831C} or deep $H_{\alpha}$imaging \citep{2019ApJ...886...74M}.   Therefore, Figure~\ref{fig:eta_mass} should be interpreted as a comparison of the characteristic magnitude of the mass-loading factor, rather than as a one-to-one comparison of identically defined quantities.

Generally, the inferred $\eta_{\rm wind}$ for Draco is comparable in order of magnitude to those obtained from GCE modeling of other low-mass systems (see the colored symbols in Figure~\ref{fig:eta_mass}). These values exhibit a trend of increasing mass-loading factor toward lower stellar masses, which is broadly consistent with theoretical expectations: lower-mass systems have shallower gravitational potential wells and are therefore less able to retain gas against stellar feedback. The observed trend has a slope in good agreement with the scaling relation derived from FIRE-2 cosmological simulations by \citet{2021MNRAS.508.2979P}. Moreover, the inferred value for Draco is close to the prediction of the FIRE-1 simulations \citep{2015MNRAS.454.2691M}.

Draco has little $\mathrm{HI}$ within the tidal radius  \citep{1999AJ....117.1758Y, 2000AJ....119..188Y, 2021ApJ...913...53P}, with an upper limit of $\mathrm{HI}$ mass  $< 8000 \rm M_\odot$ \citep{2003ApJ...588..326G}. Though such a low gas-mass measured at the present day does not provide a direct measurement of the historical mass-loading factor, it is qualitatively consistent with the strong cumulative gas removal required by our model.

\subsection{Star Formation History}
\begin{figure*}[!htb]
\includegraphics[width=0.95\textwidth,trim=0.1cm 0.1cm 0.1cm 0.1cm,clip]{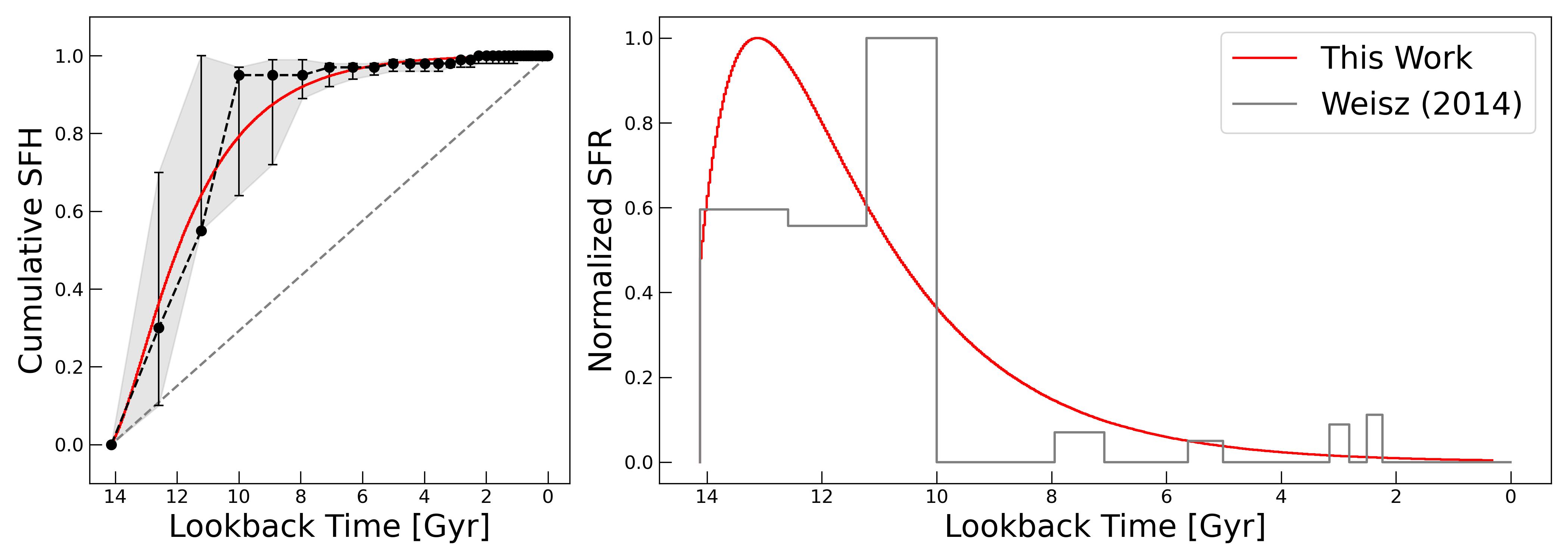}
\caption{Cumulative star formation histories (left panel) and normalized star formation rates (right panel) as a function of look-back time. The star formation history derived from CMD fitting by \citet{2014ApJ...789..148W} is shown in gray, with the shaded region indicating its uncertainty. The best-fit GCE model for the full sample is shown in red solid line}
\label{fig:sfh} 
\end{figure*}
The GCE modeling results an estimation of the SFR (Equation~\ref{func:sfr}) at each epoch. Figure~\ref{fig:sfh} presents the corresponding cumulative SFH and normalized SFR as functions of lookback time, together with observational constraints from the CMD-fitting analysis of \citet{2014ApJ...789..148W}. To facilitate comparison, all SFRs are normalized to their peak values.

For the full sample, the inferred time at which Draco assembled half of its present-day stellar mass, $\tau_{50}$, is approximately 12.0 Gyr. Accounting for the uncertainties, the cumulative SFH is broadly consistent with the CMD-fitting results of \citet{2014ApJ...789..148W}. 

The best-fit model for the full sample predicts an SFR peak approximately 1.0 Gyr after the onset of star formation, which is assumed to have occurred 14 Gyr ago. The corresponding maximum SFR is $\sim 7.61 \times10^{-5}~M_\odot{\rm yr}^{-1}$. This value is consistent with the CMD-based estimate of \citet{2001AJ....122.2524A}, but is approximately a twenty-fifth of that inferred from the chemical evolution model of \citet{2011ApJ...727...79K}. A comparison of the inferred SFHs indicates that our model favors a more extended period of star formation in Draco than the scenario proposed by \citet{2011ApJ...727...79K}.

Several factors may contribute to the discrepancy, including differences in the treatment of gas inflows and outflows, the adopted star formation law, nucleosynthetic yield prescriptions for CCSNe, SNe Ia, and AGB stars, and the assumed high-mass IMF slope. However, we suspect that the dominant source of the difference is the underlying abundance data. The stellar abundances used by \citet{2011ApJ...727...79K} are largely drawn from the same observational dataset later reanalyzed by \citet{2025ApJ...983..117H}, whose [Mg/Fe] measurements are systematically higher than those derived from APOGEE and the DESI \texttt{DD-Payne} analysis adopted in this work. Reproducing these elevated [Mg/Fe] ratios generally requires a higher star formation rate, as more rapid enrichment by core-collapse supernovae delays the decline of [Mg/Fe] associated with Type Ia supernova enrichment.

\subsection{Completeness, selection effects, and Milky Way foreground contamination}
\begin{figure*}[!htb]
\includegraphics[width=0.95\textwidth,trim=0.1cm 0.1cm 0.1cm 0.1cm,clip]{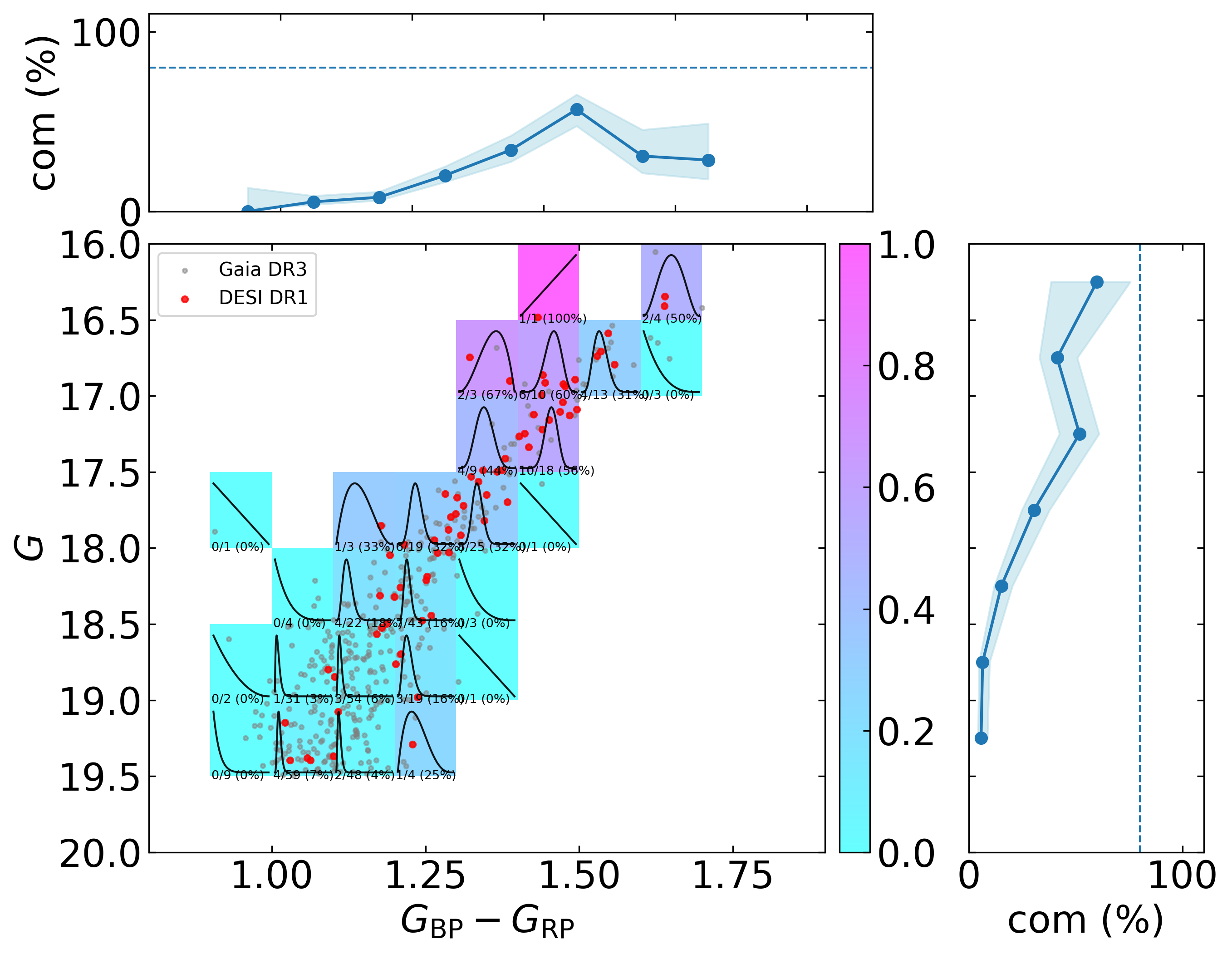}
\caption{Completeness of the DESI DR1 spectroscopic sample relative to Gaia DR3 in the CMD. The lower-left panel shows the Gaia DR3 (gray dots) and DESI DR1 (red dots) sources, with the background color scale indicating the completeness in each $(G_{\rm BP}-G_{\rm RP})$--$G$ cell. The black curves within individual cells show the posterior probability distributions of the completeness, modeled as Beta distributions based on the numbers of DESI and Gaia sources in each cell. The corresponding $N_{\rm DESI}/N_{\rm Gaia}$ values are indicated in each cell. The upper and right panels show the completeness as a function of color and $G$ magnitude, respectively. The shaded regions indicate the 16th--84th percentile credible intervals of the Beta posterior distributions, while the dashed lines mark a completeness of 80\%.}
\label{fig:completeness} 
\end{figure*}

\begin{figure*}[!htb]
\includegraphics[width=0.95\textwidth,trim=0.1cm 0.1cm 0.1cm 0.1cm,clip]{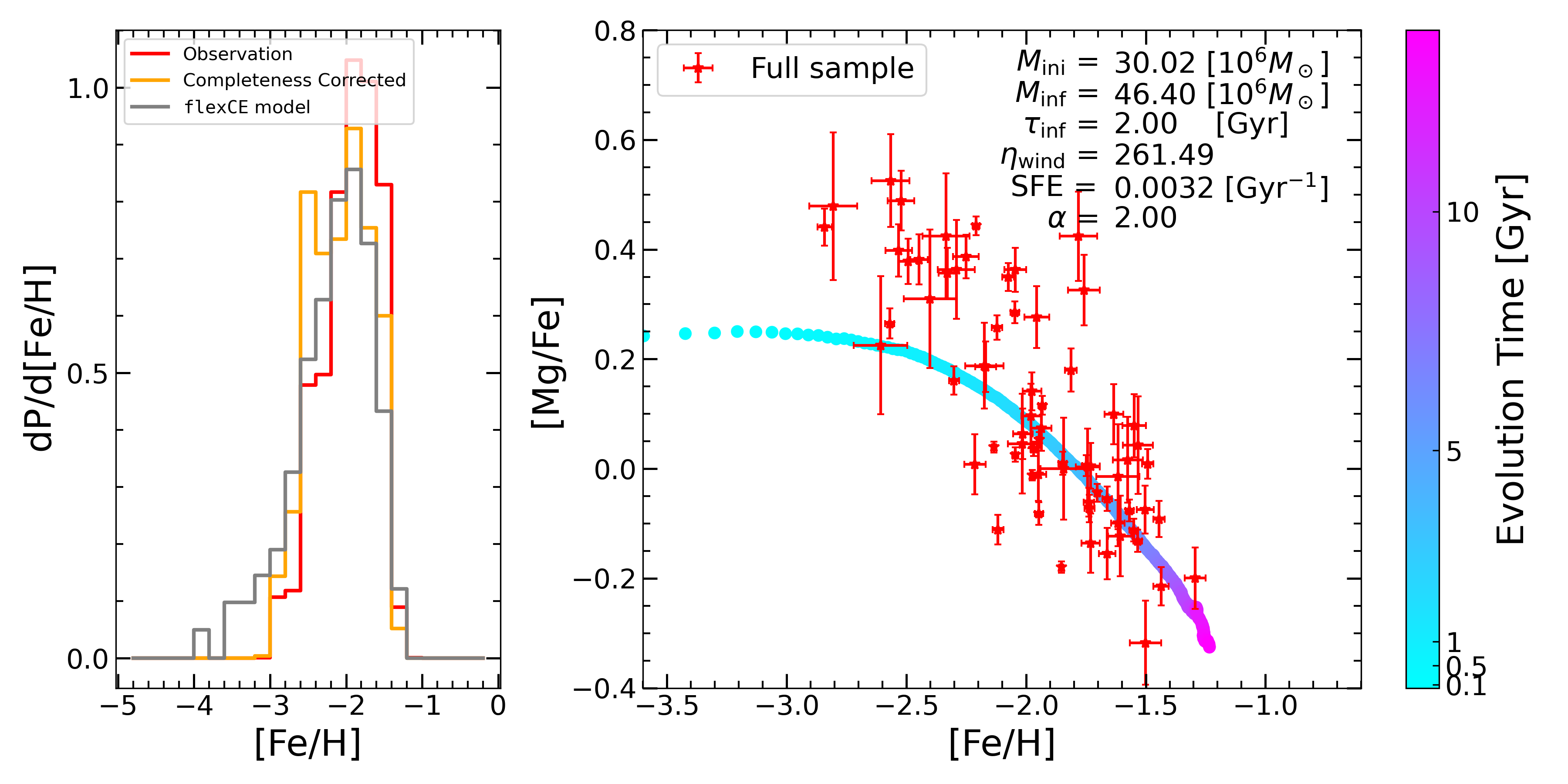}
\caption{Same as Figure~\ref{fig:draco_result}, but for the completeness-corrected sample. The left panel compares the observed MDF (red) with the completeness-corrected MDF (orange) and the best-fitting GCE model (gray). The right panel shows the $[\mathrm{Mg/Fe}]$--$[\mathrm{Fe/H}]$ distribution and the corresponding best-fitting GCE model after applying the completeness correction.}

\label{fig:draco_result_com} 
\end{figure*}

We further assess the potential impact of selection effects, sampling incompleteness, and Milky Way foreground contamination on our inferred chemical evolution path of Draco. The DESI target selection does not explicitly depends on stellar metallicity and is therefore not expected to introduce a strong metallicity-dependent selection bias. In addition, contamination from Milky Way foreground stars is mitigated by our member-star selection, which makes use of the available kinematic and astrometric information to remove stars inconsistent with Draco membership. We therefore expect the residual foreground contamination in our final spectroscopic sample to be relatively small. Nevertheless, we test explicitly whether the CMD-dependent sampling completeness can affect the inferred MDF and GCE parameters.

To estimate the sampling completeness, we use the Draco member catalog of \citet{2022ApJ...940..136P} as a reference photometric sample. We apply the same spatial, proper-motion, parallax, and brightness selection criteria used for our spectroscopic sample to this catalog, yielding 409 Draco member stars with Gaia DR3 observations. We then divide the CMD into cells with widths of 0.1 mag in $G_{\rm BP}-G_{\rm RP}$ and 0.5 mag in $G$. The sampling completeness in each CMD cell is estimated from the ratio of the number of DESI spectroscopic stars to the corresponding number of Gaia stars, 
\begin{equation}
C_{ij} = \frac{N_{{\rm DESI},ij}}{N_{{\rm Gaia},ij}},
\end{equation}
where $N_{{\rm DESI},ij}$ and $N_{{\rm Gaia},ij}$ are the numbers of DESI and Gaia stars in the $i$-th magnitude and $j$-th color bin, respectively. 

We account for the statistical uncertainty in the completeness estimates arising from the finite number of stars in each CMD cell by modeling the completeness with a beta distribution, which is naturally defined for probabilities between 0 and 1. Assuming a binomial likelihood and a uniform prior, the posterior completeness in each CMD cell is
\begin{equation}
C_{ij} \sim {\rm Beta}\left(N_{{\rm DESI},ij}+1,\ N_{{\rm Gaia},ij}-N_{{\rm DESI},ij}+1\right).
\end{equation}
This treatment naturally accounts for the statistical uncertainty in the completeness estimate arising from the finite numbers of DESI and Gaia stars in each CMD cell. The resulting completeness map is shown in Figure~\ref{fig:completeness}. The completeness varies substantially across the CMD, with generally higher sampling fractions in the brighter regions and lower completeness toward the faint end. Some individual CMD cells have particularly low completeness, reflecting the limited number of DESI targets in those regions. 

We propagate the completeness uncertainty into the MDF uncertainty analysis. For each realization, we simultaneously perturb the measured $[\mathrm{Fe/H}]$ values according to their observational uncertainties and draw a completeness value for each star from the corresponding Beta posterior. Each star is then assigned a completeness-correction weight, $w_i = \frac{1}{C_i}$, where $C_i$ is the completeness of the CMD cell containing the star. The resulting completeness-corrected MDF is then used in the GCE analysis. Figure~\ref{fig:draco_result_com} compares the original and completeness-corrected MDFs and shows the corresponding best-fitting GCE model. The completeness correction produces a modest redistribution of stars across the MDF, including a somewhat enhanced contribution from the metal-poor population. However, the overall shape and metallicity range of the MDF remain broadly consistent with those of the original sample. We therefore find no evidence that the main features of the Draco MDF are dominated by the CMD-dependent sampling of the DESI spectroscopic survey. 

We further refit the completeness-corrected MDF and $[\mathrm{Mg/Fe}]-[\mathrm{Fe/H}]$ distribution using the same GCE procedure described in Section~\ref{sec:cem}. The best-fitting model has a mass-loading factor of $\eta_{\rm wind} = 261.49^{+6.19}_{-6.45}$, and a star formation efficiency of ${\rm SFE} = 0.0032^{+0.0004}_{-0.0001}\ {\rm Gyr}^{-1}$. Although these values differ quantitatively from the corresponding parameters obtained without the completeness correction, they remain within the same general range as the values inferred from the C-normal and C-depleted subsamples. Thus, the correction modifies the detailed normalization of the inferred GCE parameters but does not qualitatively change the chemical-evolution history inferred for Draco. 

Finally, we consider the possible impact of residual Milky Way foreground contamination. Our spectroscopic sample has already been cleaned using the available kinematic and astrometric membership information, substantially reducing contamination from Galactic field stars. Moreover, the coherent $[\mathrm{Mg/Fe}]-[\mathrm{Fe/H}]$ sequence and the well-defined MDF of the selected sample indicate that the dominant population is chemically consistent with Draco. Residual foreground contamination may nevertheless contribute to the scatter of individual abundance measurements and to the tails of the MDF. Given the consistency between the original and completeness-corrected analyses, we conclude that neither CMD-dependent sampling incompleteness nor plausible residual Milky Way contamination is likely to drive the principal GCE trends or alter our main conclusions about the inefficient star formation and strong outflow-driven chemical evolution of Draco.

\section{Summary}\label{sec:sum}
Nearby dwarf galaxies provide valuable fossil records of the star formation and chemical enrichment histories of the local Universe. In this work, we investigate the chemical evolution of the Draco dwarf spheroidal galaxy using abundance measurements for 69 member stars from DESI DR1. Stellar atmospheric parameters and elemental abundances are derived with \texttt{DD-Payne}, a physically informed, data-driven machine-learning framework \citep{2019ApJS..245...34X}. By taking common stars between DESI DR1 and APOGEE DR17 as well as between DESI DR1 and LAMOST DR9 are adopted as the training set, the method allows the determination of stellar atmospheric parameters and abundances for 15 elements from the DESI spectra \citep{2024ApJS..273...19Z}.  

For accuracy, we split the DESI stellar sample of Draco into sub-populations based on their carbon abundances: C-depleted stars are mainly tip-RGB stars whose surface C-abundances were altered by the first dredge-up process but [Mg/Fe] are good fossils for galactic archaeology; C-normal stars are either early AGB stars or RGB stars that not experienced significant carbon depletion. Some of the upper-AGB stars may even experienced significant C-enhancement, and their surface [Mg/Fe] may not be perfect fossils for galactic archaeology. 

We simultaneously fit the observed MDFs and chemical abundances trend in the \mgfe-\feh planes using the GCE model tool \flexce. The model fitting returns five parameters: the slope of IMF \imf, star formation efficiency \sfe, the mass-loading factor of wind $\eta_{\rm wind}$, the inflow gas mass \ifm and inflow timescale \ift. While some of the parameters in particular the \imf, is imperfectly constrained because it is very sensitive to data quality, the star formation efficiency \sfe and the outflow mass-loading factor $\eta_{\rm wind}$ are relatively well constrained.  

The GCE modeling robustly constrains Draco to have a SFE of 0.003--0.007~${\rm Gyr}^{-1}$, the range representing systematic uncertainty. This value is comparable to recent estimates for ultra-faint dwarf galaxies. In contrast, several earlier studies reported SFEs that are approximately a factor of five higher, likely due to imperfections in the model assumptions. By combining our measurement with recent literature estimates for dwarf galaxies and stellar remnants, we derive an empirical relation between SFE and stellar mass, characterized by a power-law slope of 0.24. The GCE modeling yields an outflow mass-loading factor of $\sim$ 172--260. This aligns with the overall literature trend but deviate from earlier studies that artificially capped the loading factor at $\eta_{\rm wind} <15$.

Our results demonstrate that stellar parameters and chemical abundances measured from DESI DR1 low-resolution spectra are capable of delivering key information of star formation history of nearby dwarf galaxies. This approach is scalable to more nearby galaxies, and eventually, provides new insights into the star formation and chemical enrichment history of the local Universe. 

\begin{acknowledgments}
We gratefully thank the anonymous referee for the insightful comments and constructive suggestions. We acknowledge support from the National Natural Science Foundation of China (NSFC) under grants No.2022000083 and No.12203064. Y.H. acknowledges support from the National Key Research and Development Program of China (grant No. 2025YFF0510603). M.X. acknowledges support from the National Key Research and Development Program of China (grant No. 2022YFF0504200). C.L. acknowledges support from the National Key Research and Development Program of China (grants No. 2025YFF0511001).

The DESI Legacy Imaging Surveys consist of three individual and complementary projects: the Dark Energy Camera Legacy Survey (DECaLS), the Beijing-Arizona Sky Survey (BASS), and the Mayall z-band Legacy Survey (MzLS). DECaLS, BASS and MzLS together include data obtained, respectively, at the Blanco telescope, Cerro Tololo Inter-American Observatory, NSF’s NOIRLab; the Bok telescope, Steward Observatory, University of Arizona; and the Mayall telescope, Kitt Peak National Observatory, NOIRLab. NOIRLab is operated by the Association of Universities for Research in Astronomy (AURA) under a cooperative agreement with the National Science Foundation. Pipeline processing and analyses of the data were supported by NOIRLab and the Lawrence Berkeley National Laboratory (LBNL). Legacy Surveys also uses data products from the Near-Earth Object Wide-field Infrared Survey Explorer (NEOWISE), a project of the Jet Propulsion Laboratory/California Institute of Technology, funded by the National Aeronautics and Space Administration. Legacy Surveys was supported by: the Director, Office of Science, Office of High Energy Physics of the U.S. Department of Energy; the National Energy Research Scientific Computing Center, a DOE Office of Science User Facility; the U.S. National Science Foundation, Division of Astronomical Sciences; the National Astronomical Observatories of China, the Chinese Academy of Sciences and the Chinese National Natural Science Foundation. LBNL is managed by the Regents of the University of California under contract to the U.S. Department of Energy. The complete acknowledgments can be found at https://www.legacysurvey.org/.

\software{flexCE \citep{2017ApJ...835..224A}, emcee \citep{2013PASP..125..306F}, NumPy \citep{harris2020array}, pandas  \citep{reback2020pandas}, astropy \citep{2022ApJ...935..167A}, Matplotlib \citep{2007CSE.....9...90H}, corner \citep{Foreman-Mackey2016}
}
\end{acknowledgments}

\bibliography{sample631}{}
\bibliographystyle{aasjournal}

\end{document}
